\documentclass[11pt, a4paper]{article}
\pdfoutput=1 

\usepackage[height=8.85in,width=6.75in]{geometry}

\usepackage{graphicx,rotating}     

\usepackage{amsmath,amssymb}
\usepackage{slashed}
\usepackage{bm}
\usepackage{bbm}
\usepackage{cite}
\usepackage{comment}
\usepackage[utf8]{inputenc}
\usepackage{accents}
\usepackage[footnotesize]{caption}
\usepackage{cancel}

\newcommand{\vev}[1]{ \left\langle {#1} \right\rangle }
\newcommand{\abs}[1]{\left\vert {#1} \right\vert}

\newcommand{\hyphen}{\,\mathchar`-\mathchar`-\,}

\makeatletter
\@addtoreset{equation}{section}

\makeatother

\usepackage{xcolor}
\definecolor{darkred}{rgb}{0.7, 0., 0.}
\definecolor{orangered}{rgb}{1,0.27,0.}
\definecolor{steelblue}{rgb}{0.275,0.51, 0.706}
\definecolor{forestgreen}{rgb}{0.13,0.55,0.13}
\definecolor{tabblue}{rgb}{0.12156862745098039, 0.4666666666666667, 0.7058823529411765}

\usepackage{hyperref}
\hypersetup{colorlinks=true, linkcolor={darkred}, citecolor={tabblue}, urlcolor={tabblue}}

\usepackage{tikz}
\usepackage{tikz-feynman}
\tikzfeynmanset{compat=1.0.0}
\pgfdeclarelayer{bg}    
\pgfsetlayers{bg,main}  

\begin{document}

\hypersetup{pageanchor=false}
\begin{titlepage}

\begin{center}

\hfill IFIC/23-45 \\
\hfill FTUV-23-1005.0503\\
\hfill UMN-TH-4226/23 \\
\hfill FTPI-MINN-23-18 \\
\hfill CERN-TH-2023-186

\vskip 0.5in

{\huge \bfseries
Perturbatively including inhomogeneities \vspace{5mm}  \\
in axion inflation
} \\
\vskip .8in

{\Large Valerie Domcke$^{a}$, Yohei Ema$^{b,c}$,  and Stefan Sandner$^{d}$}

\vskip .3in
\begin{tabular}{ll}
$^a$& \!\!\!\!\!\emph{Theoretical Physics Department, CERN, 1211 Geneva 23, Switzerland}\\
$^b$& \!\!\!\!\!\emph{William I. Fine Theoretical Physics Institute, School of Physics and Astronomy, }\\[-.2em]
& \!\!\!\!\!\emph{University of Minnesota, Minneapolis, MN 55455, USA} \\
$^c$&  \!\!\!\!\!\emph{School of Physics and Astronomy, University of Minnesota, Minneapolis, MN 55455, USA} \\
$^d$& \!\!\!\!\!\emph{Instituto de F\'{\i}sica Corpuscular, Universitat de Val\`encia and CSIC,} \\[-.2em]
& \!\!\!\!\!\emph{Carrer del Catedr\'atic Jos\'e Beltr\'an Martinez 2, 46980 Paterna, Spain}
\end{tabular}

\end{center}
\vskip .6in

\begin{abstract}
\noindent
Axion inflation, i.e.\ an axion-like inflaton coupled to an Abelian gauge field through a Chern-Simons interaction, comes with a rich and testable phenomenology. 
This is particularly true in the strong backreaction regime, where the gauge field production heavily impacts the axion dynamics.
Lattice simulations have recently demonstrated the importance of accounting for inhomogeneities of the axion field in this regime. We propose a perturbative scheme to account for these inhomogeneities while maintaining high computational efficiency. 
Our goal is to accurately capture deviations from the homogeneous axion field approximation within the perturbative regime as well as self-consistently determine the onset of the non-perturbative regime.

\end{abstract}

\end{titlepage}

{
\hypersetup{linkcolor=black}
\tableofcontents
}
\renewcommand{\thepage}{\arabic{page}}
\hypersetup{pageanchor=true}

\section{Introduction}
\label{sec:introduction}

Cosmic inflation remains the most attractive theory to explain the precise observations of the cosmic microwave background (CMB) by the Planck satellite~\cite{Planck:2018jri, Planck:2018vyg}.
Among the particle physics models which can lead to a quasi-exponential expansion, considerable attention has been paid to axion-like particles as the driving field of inflation.
The axion's angular symmetry is only broken by non-perturbative effects and thus the observationally required flatness of the potential can be ensured naturally~\cite{Freese:1990rb}.
Furthermore, the shift symmetry allows for derivative couplings of the axion to the Chern-Simons density $F_{\mu\nu} \Tilde{F}^{\mu\nu}$ of a (dark) gauge field $A_\mu$.
Such couplings can lead to the exponential production of $A_\mu$ due to a tachyonic instability of one of the two helicity modes in the equations of motion (EOM), which is solely controlled by the axion velocity and hence generically impacts the later stages of inflation, corresponding to length scales much smaller than those accessible in CMB observations.
The consequences of such a large non-thermal $A_\mu$ population are diverse and include: 
i) an effective friction term in the axion EOM~\cite{Anber:2009ua}, ii) a strong enhancement of the scalar and tensor perturbations with possible observational consequences, such as the production of primordial black holes~\cite{Linde:2012bt,Bugaev:2013fya,Cheng:2015oqa,Garcia-Bellido:2016dkw,Domcke:2017fix,Garcia-Bellido:2017aan,Cheng:2018yyr} and (chiral) stochastic gravitation waves~\cite{Sorbo:2011rz,Cook:2011hg,Barnaby:2011qe, Barnaby:2011vw, Anber:2012du, Domcke:2016bkh, Garcia-Bellido:2023ser}
and iii) a mechanism for magnetogenesis~\cite{Garretson:1992vt,Anber:2006xt,Caprini:2014mja,Adshead:2016iae,Jimenez:2017cdr,Durrer:2023rhc} and baryogenesis~\cite{Anber:2015yca,Domcke:2019mnd,Domcke:2022kfs,Jimenez:2017cdr} if the gauge field is taken to be the Standard Model (SM) hypercharge. 

Obtaining accurate predictions for any of these processes requires evolving the highly non-linear system containing the axion and gauge fields, as well as when present, any light fermions. In this paper we shall focus on the case where the gauge field is a dark photon, with no other couplings to the SM (or beyond) other than the Chern-Simons coupling to the axion.\footnote{For the discussion of the SM case including light fermions see~\cite{Domcke:2018eki,Domcke:2019qmm,Gorbar:2021rlt}.}
In this case, the most important backreaction to consider is the effective friction induced by the gauge fields on the axion~\cite{Anber:2009ua}, and our interest will be in the regime where this backreaction is strong, i.e.\ typically towards the end of inflation. Changes in the axion velocity impact gauge field modes within the tachyonic instability window, which contribute to the friction force. As a result, the friction term reacts with some time delay to the changes in the axion velocity, leading to a resonantly coupled system with distinct peaks in the axion velocity~\cite{DallAgata:2019yrr,Domcke:2020zez}. These results have been confirmed in perturbative stability analysis~\cite{Peloso:2022ovc,vonEckardstein:2023gwk} as well as using the gradient expansion formalism (GEF)~\cite{Gorbar:2021rlt,vonEckardstein:2023gwk}. The latter provides a remarkably efficient tool for numerical simulations based on expressing the non-linear EOMs in position space as a tower of ordinary differential equations (ODEs) for the $2$-point functions of the axion, the gauge fields, and the gradients of the gauge fields, reducing the computation time by orders of magnitude compared to e.g.\ the iterative procedure used in~\cite{Domcke:2020zez}. 

All these methods, however, make a crucial assumption: they take the axion field to be homogeneous. Given the rapid growth of the axion perturbations, the significant departure from the standard slow-roll regime and the strong non-linearities involved, it is not surprising that this approximation breaks down in the strong backreaction regime. This has recently been explicitly demonstrated in a lattice simulation~\cite{Figueroa:2023oxc} using \texttt{CosmoLattice}~\cite{Figueroa:2020rrl,Figueroa:2021yhd}, reproducing earlier results when switching off axion gradients but finding significant departures when the axion gradients are taken into account consistently. While accurately dealing with the full non-linear problem including the strong backreaction, the downside of these simulations is that they are extremely costly. Moreover, as can be seen from the results obtained in~\cite{Figueroa:2023oxc}, the highly non-linear dynamics implies that observable quantities (such as the magnitude of the gauge field contribution or the scalar perturbations at a given scale) are not monotonic functions of the axion gauge field coupling.
A full exploration of the phenomenology of axion inflation throughout the parameter space thus seems very costly at best based on lattice simulations only. 

Our goal in this paper is to leverage the benefits of the highly efficient GEF to perturbatively include axion gradients. To obtain a closed set of ODEs, this requires evolving not only the $2$-point functions but also higher $p$-point functions under the GEF scheme. We are particularly interested in the regime where axion gradients are relevant and impact the evolution of the axion vacuum expectation value and the gauge field distribution, while still allowing for a perturbative treatment. As time evolves, this will typically give way to a regime in which the axion gradients become too large to be treated perturbatively, calling for a lattice simulation. Within our formalism, we self-consistently determine the breakdown of perturbativity, providing a tool to compute initial conditions for lattice simulations, focusing their computational power on the truly non-linear regimes. Our work should be seen as a first step in developing this methodology, and we discuss possible extensions and scalability.
In the process, we gain new insights into the application of the GEF to axion inflation, notably regarding the need to go to rather high order in the GEF tower to obtain convergence as well as an improved truncation relation.

The paper is organized as follows.
In Sec.~\ref{sec:gef} we briefly review the GEF and derive an extension including axion gradients.
Some technical details, in particular the more lengthy equations for the 3-point functions are given in App.~\ref{app:eom_3pt}, while App.~\ref{app:trunc} focuses on the limitations of the GEF and in particular the truncation relation. 
Our results are presented in Sec.~\ref{sec:num}, and contrasted with results obtained in lattice computations
as well as under the assumption of a homogeneous axion field.    
The final Sec.~\ref{sec:conclusion} summarizes and discusses the results.

\addtocontents{toc}{\vspace{-0.5em}} 
\section{The gradient expansion formalism including axion inhomogeneities}
\label{sec:gef}

The gradient expansion formalism developed in Refs.~\cite{Gorbar:2021rlt,Gorbar:2021zlr} for axion inflation (see \cite{Sobol:2019xls,Sobol:2020lec} for earlier work in related contexts) provides a computationally efficient way of accounting for the backreaction of the gauge fields on the axion dynamics. The system of interest here is an unbroken, dark $U(1)$ gauge group coupled to the axion via a Chern-Simons interaction,
\begin{align}
	S &= \int d^4x \sqrt{-g}\left[\frac{1}{2}g^{\mu\nu}\partial_\mu \phi \partial_\nu \phi - V(\phi)
	- \frac{1}{4}g^{\mu\rho}g^{\nu\sigma}F_{\mu\nu}F_{\rho\sigma}
	- \frac{\alpha \phi}{4\pi f_a}\frac{1}{\sqrt{-g}} F_{\mu\nu}\tilde{F}^{\mu\nu}\right]\,,
	\label{eq:action}
\end{align}
where $\tilde{F}^{\mu\nu} = \epsilon^{\mu\nu\rho\sigma}F_{\rho\sigma}/2$ with $\epsilon^{0123} = 1$ and for simplicity we will set $V(\phi) = m_\phi^2 \phi^2/2$.
The GEF re-arranges the resulting coupled partial differential equations governing the dynamics of the electric and magnetic field (i.e.\ Maxwell's equations in the presence of an axion photon coupling) into a tower of linear ODEs for the $2$-point functions
\begin{align}
	\mathcal{P}_{X}^{(n)} &= \frac{1}{a^n}\left\langle \vec{X}\cdot(\vec{\nabla}\times)^n \vec{X}\right\rangle\,,
	\quad
	\mathcal{P}_{XY}^{(n)} = -\frac{1}{a^n}\left\langle \vec{X}\cdot(\vec{\nabla}\times)^n \vec{Y}\right\rangle\,.
	\label{eq:2pt}
\end{align}
with $X,Y = \{E,B\}$, the bracket indicating the spatial average, 
and $a$ indicating the scale factor of the expanding universe.
Under the assumptions that the axion field is homogeneous and that its velocity varies only slowly, 
the EOMs of the $2$-point functions form a closed system, and
the infinite set of equations can be truncated at finite power $n$ of the curl yielding an efficient and accurate evaluation of the axion and gauge field dynamics~\cite{Gorbar:2021rlt}. However, as we will discuss in more detail below, typically neither of these assumptions are fulfilled once the gauge field backreaction becomes significant. In the strong backreaction regime, the axion field enters into a oscillatory regime, understood as resonances resulting from the time-delayed friction force exerted by the gauge fields~\cite{DallAgata:2019yrr,Domcke:2020zez,Peloso:2022ovc,Gorbar:2021rlt}. 
More recently lattice studies have qualitatively confirmed the existence of the oscillatory behavior, finding however quantitatively significant differences (notably a significant damping of these oscillations) when including the inhomogeneities in the axion field~\cite{Figueroa:2023oxc}. 

The goal of the present paper is to extend the GEF beyond these two assumptions. Below, we derive an extended version of the GEF including axion inhomogeneities (i.e.\ axion gradient terms) in a perturbative manner. The resulting non-linearities in the EOMs of the $2$-point functions prompt us to include higher $p$-point functions in order to reduce our equations to ODEs, in a similar spirit as the original GEF. Similarly, we will need to find a suitable procedure to truncate this second expansion series as finite (and low) $p$. In the process, we will shed light on the role of rapid changes in the axion velocity and the resulting limitations of the GEF.

\subsection{Equations of motion}

We start from the exact EOMs as derived from Eq.~\eqref{eq:action}, separating the homogeneous component of the axion $\phi(t)$ from its inhomogeneous component $\chi(t,\vec x)$,
\begin{align}
	0 &= \ddot{\phi} + 3H \dot{\phi} + m_\phi^2 \phi
	- \frac{\beta}{M_P}\vev{\vec{E}\cdot\vec{B}}\,, \label{eq:eom_phi}
	\\
	0 &= \ddot{\chi} + 3H \dot{\chi} - \frac{\nabla^2 \chi}{a^2} + m_\phi^2 \chi
	- \frac{\beta}{M_P}\left(\vec{E}\cdot\vec{B} - \vev{\vec{E}\cdot \vec{B}}\right)\,, \label{eq:eom_chi}
	\\
	0 &= \dot{\vec{E}} + 2 H \vec{E}- \frac{1}{a}\vec{\nabla}\times \vec{B}
	+ \frac{\beta}{M_P}\left(\dot{\phi} + \dot{\chi}\right) \vec{B}
	+ \frac{\beta}{M_P}\frac{1}{a}\vec{\nabla}\chi \times \vec{E}\,, \label{eq:eom_E}
	\\
	0 &= \dot{\vec{B}} + 2H \vec{B} + \frac{1}{a}\vec{\nabla}\times \vec{{E}}\,, \label{eq:eom_B}
	\\	
	0 &= \vec{\nabla}\cdot \vec{E}
	+ \frac{\beta}{M_P} \vec{\nabla}\chi \cdot \vec{B}\,,
	\quad
	0 = \vec{\nabla}\cdot \vec{B}\,, \label{eq:constraint}
\end{align}
for the matter sector with $\beta = \alpha M_P/(\pi f_a)$, and
\begin{align}
	H^2 &= \frac{1}{3M_P^2}\left\langle\frac{1}{2}\left(\dot{\phi}^2 + \dot{\chi}^2\right) 
	+ \frac{(\partial_i \chi)^2}{2a^2} + \frac{m_\phi^2}{2}\left(\phi^2 + \chi^2\right)
	+ \frac{1}{2}\left(\vert{\vec{E}}\vert^2 + \vert{\vec{B}}\vert^2\right)
	\right\rangle\,,
	\\
    \label{eq:eom_H}
	\dot{H} &= - \frac{1}{6M_P^2}\left\langle3\left(\dot{\phi}^2 + \dot{\chi}^2\right) 
	+ \frac{(\partial_i \chi)^2}{a^2} 
	+ 2  \left(\vert{\vec{E}}\vert^2 + \vert{\vec{B}}\vert^2\right)\right\rangle\,,
\end{align}
for the gravity sector.
For $\chi = 0$, multiplying Eqs.~\eqref{eq:eom_E} and~\eqref{eq:eom_B} with $\{(\vec \nabla \times)^n \vec E, (\vec \nabla \times)^n \vec B \}$ yields a tower of ODEs linear in the (scalar) $2$-point functions~\eqref{eq:2pt}.
To include the axion gradients, the structure of the last two terms in Eq.~\eqref{eq:eom_E} as well as the last term in the Gauss constraint in Eq.~\eqref{eq:constraint} suggests the inclusion of $3$-point functions including one power of either $\dot \chi$ or $\vec \nabla \chi$. Using the EOMs, the ODEs governing these 3-point functions in turn depend on $4$-point functions, etc. The result is a double expansion in gradients $(\nabla \times)^n$ and $p$-point functions.

To first order in this expansion, we will only keep $3$-point functions with up to one spatial derivative.
In this case, the EOMs of the $2$-point electromagnetic functions ${\cal P}$ with $n \geq 2$ are the same as in the usual GEF formalism~\cite{Gorbar:2021rlt},
\begin{align}
    \label{eq:eom_PEn}
	&\dot{\mathcal{P}}_E^{(n)} + (n+4)H \mathcal{P}_E^{(n)} 
	- \frac{2\beta \dot{\phi}}{M_P} \mathcal{P}_{EB}^{(n)} 
	+ 2\mathcal{P}_{EB}^{(n+1)}
	= \left[\dot{\mathcal{P}}_E^{(n)}\right]_{b}\,, \\
    \label{eq:eom_PBn}
	&\dot{\mathcal{P}}_B^{(n)} + (n+4)H\mathcal{P}_B^{(n)} - 2 \mathcal{P}_{EB}^{(n+1)}
	= \left[\dot{\mathcal{P}}_B^{(n)}\right]_{b}\,, \\
    \label{eq:eom_PEBn}
	&\dot{\mathcal{P}}_{EB}^{(n)} + (n+4)H \mathcal{P}_{EB}^{(n)}
	- \mathcal{P}_E^{(n+1)} + \mathcal{P}_B^{(n+1)} - \frac{\beta \dot{\phi}}{M_P} \mathcal{P}_{B}^{(n)}
	= \left[\dot{\mathcal{P}}_{EB}^{(n)}\right]_{b}\,.
\end{align}
Here, the boundary terms on the right-hand side account for the change in the number of modes which have been excited from the vacuum, see below for a more detailed discussion.
The superscript $(n)$ refers to the number of curls, indexing the GEF tower.

The EOMs for $2$-point functions for the electromagnetic fields for $n = \{0,1\}$ now contain $3$-point functions ${\cal B}$ as anticipated,
\begin{align}
    \label{eq:PE0}
	&\dot{\mathcal{P}}_E^{(0)} + 4H\mathcal{P}_E^{(0)} + 2 \mathcal{P}_{EB}^{(1)} 
	- \frac{2\beta \dot{\phi}}{M_P} \mathcal{P}_{EB}^{(0)} - \frac{2\beta}{M_P}\mathcal{B}_{\dot{\chi};EB}^{(0)}
	= \left[\dot{\mathcal{P}}_E^{(0)}\right]_b\,, \\
    \label{eq:PB0}
	&\dot{\mathcal{P}}_B^{(0)} + 4H\mathcal{P}_B^{(0)} - 2 \mathcal{P}_{EB}^{(1)}
	= \left[\dot{\mathcal{P}}_B^{(0)}\right]_{b}\,, \\
    \label{eq:PEB0}
	&\dot{\mathcal{P}}_{EB}^{(0)} + 4H \mathcal{P}_{EB}^{(0)}
	- \mathcal{P}_E^{(1)} + \mathcal{P}_B^{(1)} - \frac{\beta \dot{\phi}}{M_P} \mathcal{P}_{B}^{(0)}
	- \frac{\beta}{M_P}\mathcal{B}_{\dot{\chi}; B}^{(0)}
	- \frac{\beta}{M_P}\left(\mathcal{B}_{\chi;EB}^{(1,0)} - \mathcal{B}_{\chi;EB}^{(0,1)}\right)
	= \left[\dot{\mathcal{P}}_{EB}^{(0)}\right]_{b}\,,
\end{align}
and 
\begin{align}
    \label{eq:PE1}
	&\dot{\mathcal{P}}_E^{(1)} + 5H\mathcal{P}_E^{(1)} + 2 \mathcal{P}_{EB}^{(2)} 
	- \frac{2\beta \dot{\phi}}{M_P} \mathcal{P}_{EB}^{(1)} 
	- \frac{2\beta}{M_P}\mathcal{B}_{\dot{\chi};EB}^{(1,0)}
	= \left[\dot{\mathcal{P}}_E^{(1)}\right]_b\,, \\
    \label{eq:PB1}
	&\dot{\mathcal{P}}_B^{(1)} + 5H\mathcal{P}_B^{(1)} - 2 \mathcal{P}_{EB}^{(2)}
	= \left[\dot{\mathcal{P}}_B^{(1)}\right]_{b}\,, \\
    \label{eq:PEB1}
	&\dot{\mathcal{P}}_{EB}^{(1)} + 5H \mathcal{P}_{EB}^{(1)}
	- \mathcal{P}_E^{(2)} + \mathcal{P}_B^{(2)} 
	- \frac{\beta \dot{\phi}}{M_P} \mathcal{P}_{B}^{(1)}
	- \frac{\beta}{M_P}\mathcal{B}_{\dot{\chi}; B}^{(1)}
	= \left[\dot{\mathcal{P}}_{EB}^{(1)}\right]_{b}\,,
\end{align}
where we have defined
\begin{align}
	\mathcal{B}_{f;E}^{(n)} &= \frac{1}{a^n}
	\vev{f \left(\left(\vec{\nabla}\times\right)^n \vec{E}\right)\cdot \vec{E}}\,,
	\quad
	\mathcal{B}_{f;B}^{(n)} = \frac{1}{a^n}
	\vev{f \left(\left(\vec{\nabla}\times\right)^n  \vec{B}\right)\cdot \vec{B}}\,,
	\quad
	\mathcal{B}_{f;EB}^{(0)} = -\vev{f \vec{E}\cdot \vec{B}}\,,
	\nonumber \\
	\mathcal{B}_{f;EB}^{(1,0)} &= -\frac{1}{a}\vev{f \left(\vec{\nabla}\times \vec{E}\right)\cdot \vec{B}}\,,
	\quad
	\mathcal{B}_{f;EB}^{(0,1)} = -\frac{1}{a}\vev{f \vec{E}\cdot \left(\vec{\nabla}\times\vec{B}\right)}\,,
\end{align}
with $f = \chi, \dot{\chi}$ and $n = \{ 0, 1 \}$.
The two superscripts on the 3-point functions refer to the number of curls acting on $\vec E$ and $\vec B$, respectively.
The EOMs for these 3-point functions can be obtained analogously from Eqs.~\eqref{eq:eom_E} and \eqref{eq:eom_B} and are given explicitly in App.~\ref{app:eom_3pt}. 
Importantly, we note here the key approximations which enter in their derivation: Firstly, we keep only $3$-point functions containing up to one spatial derivative. This is not a fundamental limitation of the proposed method but should rather be seen as the first order in a gradient expansion series. Including higher order terms will lead to more (and more lengthy) equations, with the number of equations scaling at most\footnote{The scaling will be milder with suitable optimizing of the equations, using in particular integration by parts. } as $n_{\cal B}^3$ with $n_{\cal B}$ the maximal number of derivatives appearing in the $3$-point function. However, the overall structure of the equations remains unchanged and we expect this to be numerically tractable. 
The results derived in this way are reliable as long as the subsequent order in this expansion is sufficiently small, and as discussed below, we will use this as a criterion to self-consistently check the validity of our method.
Secondly, we factorize the resulting $4$-point functions appearing in the EOMs of the $3$-point functions into products of $2$-point functions assuming Gaussian distributions for the electromagnetic fields.\footnote{
Note that in the presence of axion gradients, the electric field is no longer divergence free. Defining $\vec D = \vec E + (\beta / M_P) \chi \vec B$, we can rewrite our expressions in terms of a divergence free quantity, $\vec \nabla \cdot \vec D = 0$. The price to pay is the appearance of $5$-point functions in the EOMs of the $3$-point functions, whose factorization requires assumptions not only on the gaussianity of the gauge fields but also of the axion perturbation $\chi$. While for the former, this is in good agreement with results found in lattice simulations (we thank Dani Figueroa and Ander Urio Garmendia for providing valuable input and cross-checks on this point), the axion fluctuations are expected to be highly non-gaussian due to the non-linear source terms. Therefore, under the approximations we employ we find it more convenient to work with original electric field, taking into account that it is not divergence free. 
} 
We moreover set $4$-point functions that involve one factor of $\vec{\nabla}\chi$ to vanish 
since this leaves a spatial index uncontracted within the $2$-point function, which would be in violation of statistical isotropy.

Finally the $2$-point functions of the axion fluctuations evolve as 
\begin{align}
    \label{eq:Pchi0}
	&\dot{\mathcal{P}}^{(0)}_{\chi} - 2 \mathcal{P}^{(0)}_{\chi \dot{\chi}} = 0\,, \\
    \label{eq:Pchidchi0}
	&\dot{\mathcal{P}}^{(0)}_{\chi \dot{\chi}} + 3 H \mathcal{P}_{\chi \dot{\chi}}^{(0)}
	+ m_\phi^2 \mathcal{P}_\chi^{(0)} + \frac{\beta}{M_P}\mathcal{B}_{\chi;EB}^{(0)}
	- \mathcal{P}_{\dot{\chi}}^{(0)}
	= 0\,, \\
    \label{eq:Pdchi0}
	&\dot{\mathcal{P}}^{(0)}_{\dot{\chi}} + 6H\mathcal{P}^{(0)}_{\dot{\chi}} 
	+ 2 m_\phi^2 \mathcal{P}^{(0)}_{\chi\dot{\chi}} + \frac{2\beta}{M_P}\mathcal{B}_{\dot{\chi};EB}^{(0)}
	= 0\,,
\end{align}
with
\begin{align}
	\mathcal{P}_{\chi}^{(2n)} = \frac{1}{a^{2n}}\vev{\chi \nabla^{2n} \chi}\,,
	\quad
	\mathcal{P}_{\dot{\chi}}^{(2n)} = \frac{1}{a^{2n}}\vev{\dot{\chi} \nabla^{2n} \dot{\chi}}\,,
	\quad
	\mathcal{P}_{\chi \dot{\chi}}^{(2n)} =\frac{1}{a^{2n}}\vev{\chi \nabla^{2n} \dot{\chi}}\,.
\end{align}

\subsection{Boundary terms and truncation}

Two subtleties in the derivation above deserve a more detailed discussion: the boundary terms in Eqs.~\eqref{eq:PE0} to~\eqref{eq:PEB1} and the truncation of the gradient expansion series for the $2$-point functions at finite $n$. Both of these are a priori not related to the inclusion of the axion inhomogeneities as they appear only in the EOMs of the $2$-point functions. In fact, for a homogeneous axion field, these are the only two approximations by which the GEF prescription deviates from an exact solution and this is where the approximation of a slowly varying axion velocity enters. However, while for a homogeneous axion field the gauge field power spectra ${\cal P}^{(0)}_{E,B,EB}$ have been shown to be robustly reproducing the results found solving the mode equations of the gauge field~\cite{Gorbar:2021rlt} and are moreover in very good agreement with lattice results after setting the axion gradient terms to zero~\cite{Figueroa:2023oxc}, we find the impact of these approximations on the higher orders in the GEF tower ${\cal P}^{(n)}_{E,B,EB}$ to be significant (see also \cite{vonEckardstein:2023gwk}). 
We will show that these difficulties can be mitigated by an improved truncation relation. Including axion gradients, we observe that this ensures sufficient stability in the algorithm within the perturbative regime of axion inhomogeneities. 

\paragraph{Boundary terms.}
In terms of the mode functions of the vector potential $A$, the gauge field $2$-point functions are  given as
\begin{align}
\label{eq:PEn_mode_integral}
	\mathcal{P}_E^{(n)} &= \frac{1}{a^{n+4}}\int\frac{d^3 k}{(2\pi)^3}\theta(k_h(t)-k) \sum_\sigma
	(\sigma k)^n \abs{\frac{dA_\sigma}{d\tau}}^2\,,
	\\
	\mathcal{P}_B^{(n)} &= \frac{1}{a^{n+4}}\int\frac{d^3 k}{(2\pi)^3}\theta(k_h(t)-k) \sum_\sigma
	(\sigma k)^{n+2} \abs{A_\sigma}^2\,,
	\\
 \label{eq:PEBn_mode_integral}
	\mathcal{P}_{EB}^{(n)} &= \frac{1}{2a^{n+4}}\int\frac{d^3 k}{(2\pi)^3}\theta(k_h(t)-k) \sum_\sigma
	(\sigma k)^{n+1} \frac{d}{d\tau}\abs{A_\sigma}^2\,,
\end{align}
where $\sigma$ encodes the two gauge field polarizations and the Heaviside function ensures the vacuum subtraction, i.e.\ that only modes with $k < k_h$ which have been exited out of their vacuum state contribute to the regularized integral. To determine $k_h$, we refer to the EOM for $A_k(\tau)$,
\begin{align}
\label{eq:A_mode_ODE}
\frac{d^2 A_\sigma}{d\tau^2} + k\left(k - \lambda\sigma 2\xi a H\right)A_\sigma = 0\,, \quad \lambda =  \text{sign}(\dot \phi)\,,
\end{align}
which encounters a tachyonic instability for the polarization $\sigma = \lambda$ for 
\begin{align}
\label{eq:def_kh}
k_h(t) = \underset{t' \leq t}{\mathrm{max}} \left[2\xi(t') a(t') H(t')\right] \quad \text{with  } \xi = \beta |\dot \phi| / (2 H M_P)\,.
\end{align}
Here we assume that the axion inhomogeneity does not affect the gauge boson dynamics at the sub-horizon scales. 
Since the gauge bosons are excited only after exiting the horizon, 
the axion inhomogeneity produced from the gauge bosons is also expected to be outside the horizon, justifying our assumption.
Taking this into account leads to boundary terms in the EOMs for the $2$-point functions, which are explicitly given as 
\begin{align}
	\left[\dot{\mathcal{P}}_E^{(n)}\right]_{b}
	&= \frac{\dot{k}_h}{a^{n+4}}\frac{k_h^2}{2\pi^2}\sum_{\sigma}(\sigma k_h)^n 
	\abs{\frac{dA_\sigma}{d\tau}}_{k=k_h}^2\,, \label{eq:PEn}
	\\
	\left[\dot{\mathcal{P}}_B^{(n)}\right]_{b}
	&= \frac{\dot{k}_h}{a^{n+4}}\frac{k_h^2}{2\pi^2}\sum_{\sigma}(\sigma k_h)^{n+2}
	\abs{A_\sigma}_{k=k_h}^2\,,
	\\
	\left[\dot{\mathcal{P}}_{EB}^{(n)}\right]_{b}
	&= \frac{\dot{k}_h}{a^{n+4}}\frac{k_h^2}{2\pi^2}\sum_{\sigma}(\sigma k_h)^{n+1} 
	\mathrm{Re}\left[A_\sigma^* \frac{dA_\sigma}{d\tau}\right]_{k=k_h}\,. \label{eq:PEBn}
\end{align}
To evaluate these, we follow the prescription given in~\cite{Gorbar:2021rlt} which is based on the solutions for the mode functions for constant $\xi$. 
The boundary terms thus arise from the need to regularize the vacuum contribution, and in this implementation, rely on an at most slowly varying axion velocity. 
The final results are not particularly sensitive to the value of the cut-off $k_h$ since
the change of the integration range in Eqs.~\eqref{eq:PEn_mode_integral}--\eqref{eq:PEBn_mode_integral} by changing $k_h$ is compensated by the change of the boundary term.
For example, reducing $k_h$ by $25\%$ induces changes which are smaller than $10\%$ (and largely smaller than $1\%$) in $\xi$.

\paragraph{Truncation relation.} 
The infinite tower of equations~\eqref{eq:eom_PEn} to~\eqref{eq:eom_PEBn} needs to be truncated at some finite $n$ to implement the GEF numerically. The truncation relation employed in~\cite{Gorbar:2021rlt} is
\begin{align}
	\mathcal{P}_X^{(n_\mathrm{max}+1)} &\simeq \left(\frac{k_h}{a}\right)^2 \mathcal{P}_X^{(n_\mathrm{max}-1)}\,,
	\label{eq:trunc}
\end{align}
where $X = E, B$ or $EB$. This can be derived from Eqs.~\eqref{eq:PEn} to \eqref{eq:PEBn} assuming a power law spectrum for $|A_\sigma(k)|^2$ around $k_h$, as proposed in Ref.~\cite{Gorbar:2021rlt}. However, both results from lattice simulations~\cite{Figueroa:2023oxc} as well as the results obtained using solving the gauge field mode equations show that there can be a lot of structure in the spectrum at this scale, leading to significant deviations in the ratio $\mathcal{P}_X^{(n + 1)}/\mathcal{P}_X^{(n-1)}$, as shown in App.~\ref{app:trunc}. 
However, as we show in that appendix, Eq.~\eqref{eq:trunc} can also be obtained (without invoking the power law approximation of the spectrum) as the asymptotic large-$n$ limit under the assumption that $\xi$ is constant. As we show there, a value of $n \sim 55$ is necessary for the relation~\eqref{eq:trunc} to hold up to about $5\%$.
To our understanding, this explains why values of $n \sim 100$ are necessary to achieve convergence in the GEF, whereas $n \gg 1$ should have been sufficient based on the assumption of power law spectrum around $k_h$. 

This in turn indicates that rapid changes in $\xi$ will induce errors in this truncation relation, which propagate through the coupled system of equations down to low $n$. 
This is supported by our observations in the numerical studies shown in App.~\ref{app:trunc} and was also recently observed in Ref.~\cite{vonEckardstein:2023gwk}. 
When the axion velocity drops rapidly, the $\mathcal{P}_X^{(n)}$ take (unphysically) large values at large $n$, which over time propagate down to lower $n$ modes. If the phase of rapidly changing (in particular dropping) axion velocity is sufficiently long, this can in principle impact the observables, i.e.\ the  $\mathcal{P}_X^{(0)}$ power spectra. 
This calls for an improvement of the truncation relation~\eqref{eq:trunc} in the further development of the GEF formalism.
We show in App.~\ref{app:trunc} results obtained using not only $\mathcal{P}_X^{(n_\text{max} - 1)}$ but instead a series of $\mathcal{P}_X^{(n_\text{max} + 1 - 2 l)}$ to determine  $\mathcal{P}_X^{(n_\text{max} + 1)}$,
\begin{align}
\label{eq:trunc_improved}
	\bar{\mathcal{P}}_X^{(n_\mathrm{max}+1)}
	= \sum_{l = 1}^L (-1)^{l - 1} \begin{pmatrix}  L \\ l \end{pmatrix}
	 \bar{\mathcal{P}}_X^{(n_\mathrm{max}+1-2l)} \,,
\end{align}
where $\bar{\mathcal{P}}_X^{(n)} = \mathcal{P}_X^{(n)}/H_0^4 (k_h/a)^n$. For $L = 1$ we trivially recover Eq.~\eqref{eq:trunc}. As we show in App.~\ref{app:trunc} for $L = 4$ and $L = 10$ this improves the stability of the system to a point which is sufficient for the study of the regime of perturbative axion inhomogeneities.

\subsection{Validity of the axion gradient expansion}
\label{sec:consistency}
In the scheme described above we included only terms with up to one power of the axion gradient. To ensure the consistency of this approach, we monitor the second order axion derivatives sourced in this manner (without including their backreaction on the truncated system described above). More precisely, we monitor the ratio of the gradient to the kinetic energy,
\begin{align}
	R_\chi \equiv \left\vert \frac{\vev{(\nabla \chi)^2}}{\dot{\phi}^2 + \vev{\dot{\chi}^2}}\right\vert\,.
\end{align}
As long as this quantity is small, it is justified to drop higher powers of the axion gradients, whereas $R_\chi$ above ${\cal O}(0.1\hyphen1)$ indicates that the axion gradients can no longer be treated perturbatively, calling for a full lattice simulation. To compute $\vev{(\nabla \chi)^2} = -\vev{\chi \nabla^2\chi}$, the relevant EOMs for the axion $2$-point functions to second order in the gradient expansion are
\begin{align}
	&\dot{\mathcal{P}}_{\chi}^{(2)} + 2 H \mathcal{P}_{\chi}^{(2)}
	- 2\mathcal{P}_{\chi\dot{\chi}}^{(2)} = 0\,,
	\\
	&\dot{\mathcal{P}}_{\chi\dot{\chi}}^{(2)} + 5H\mathcal{P}_{\chi\dot{\chi}}^{(2)}
	+ m_\phi^2 \mathcal{P}_{\chi}^{(2)}
	+ \frac{\beta}{M_P}\mathcal{B}_{\chi; EB}^{(2; 0,0)} - \mathcal{P}_{\dot{\chi}}^{(2)} = 0\,,
	\\
	&\dot{\mathcal{P}}_{\dot{\chi}}^{(2)}
	+ 8H\mathcal{P}_{\dot{\chi}}^{(2)} + 2m_\phi^2 \mathcal{P}_{\chi\dot{\chi}}^{(2)}
	+ \frac{2\beta}{M_P}\mathcal{B}_{\dot{\chi}; EB}^{(2; 0,0)} 
	= 0\,,
\end{align}
which in turn require the evaluation of the 3-point function $\mathcal{B}_{f;EB}^{(2;0,0)}$, whose definitions and equations are again given in the App.~\ref{app:eom_3pt}.
The breakdown of this perturbative expansion scheme $R_\chi > 0.5$ is indicated by the dark gray region in the figures below.

\section{Numerical results}
\label{sec:num}

\begin{figure}[!t]
	\centering
 	\includegraphics[width=0.32\linewidth]{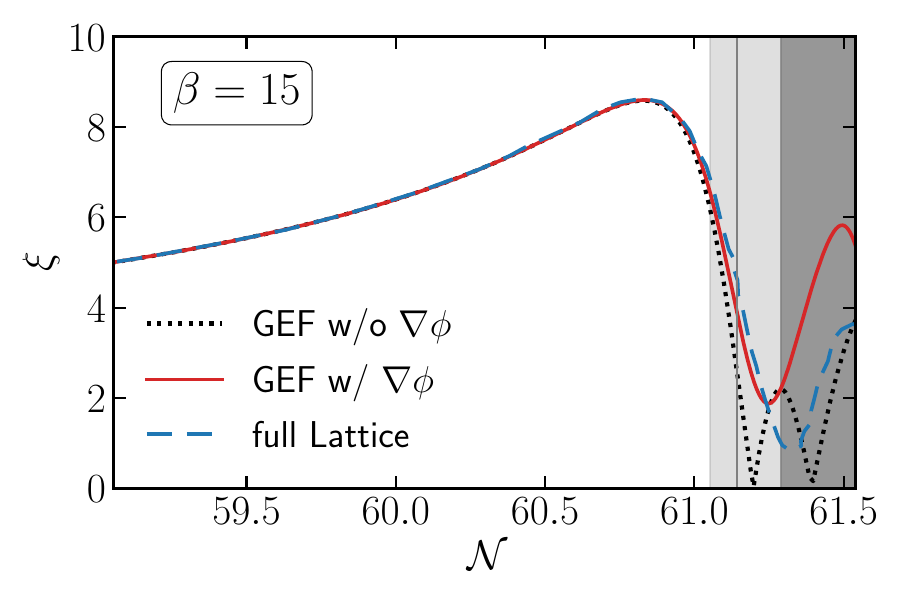}
	\hspace{0.5mm}
 	\includegraphics[width=0.32\linewidth]{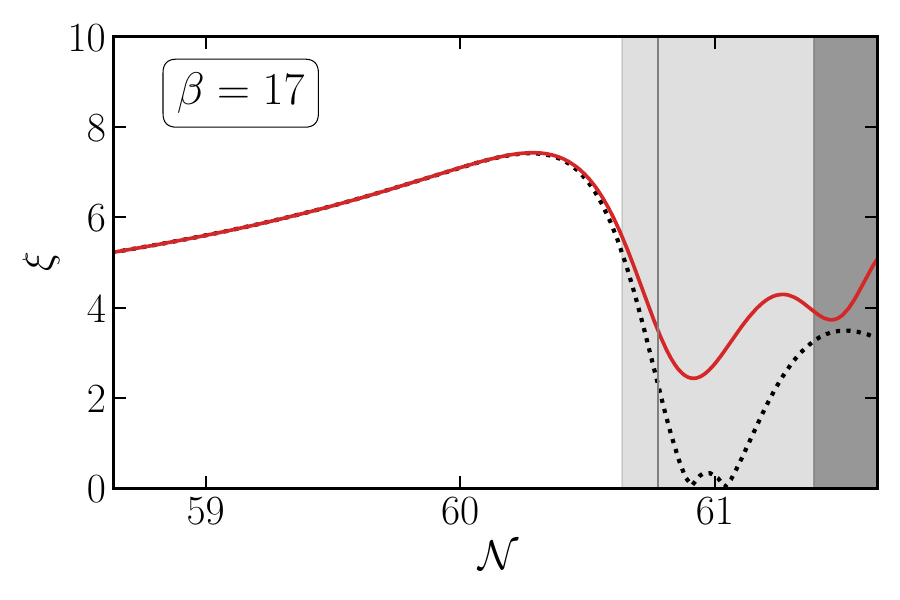}
	\hspace{0.5mm}
	\includegraphics[width=0.32\linewidth]{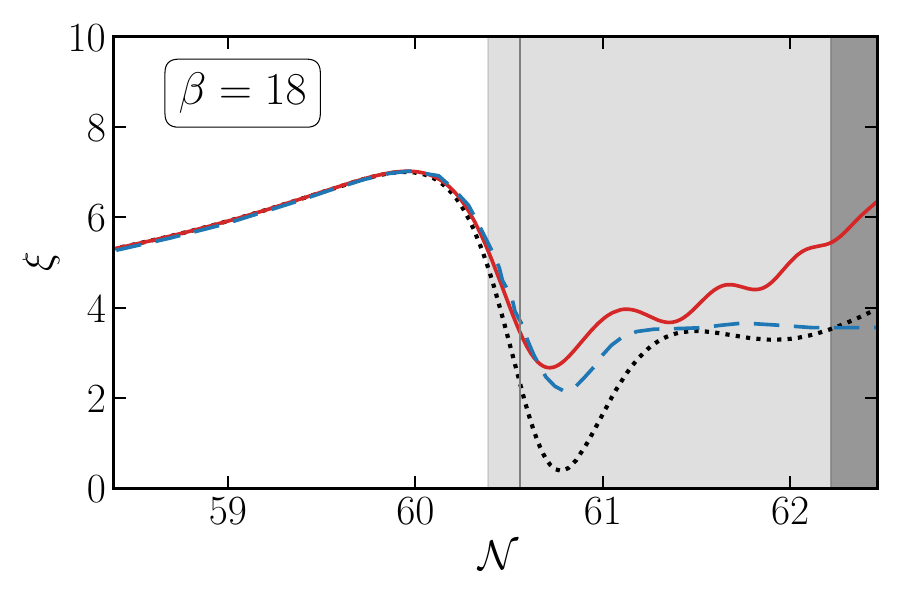}
    \vspace{1mm}
	\includegraphics[width=0.32\linewidth]{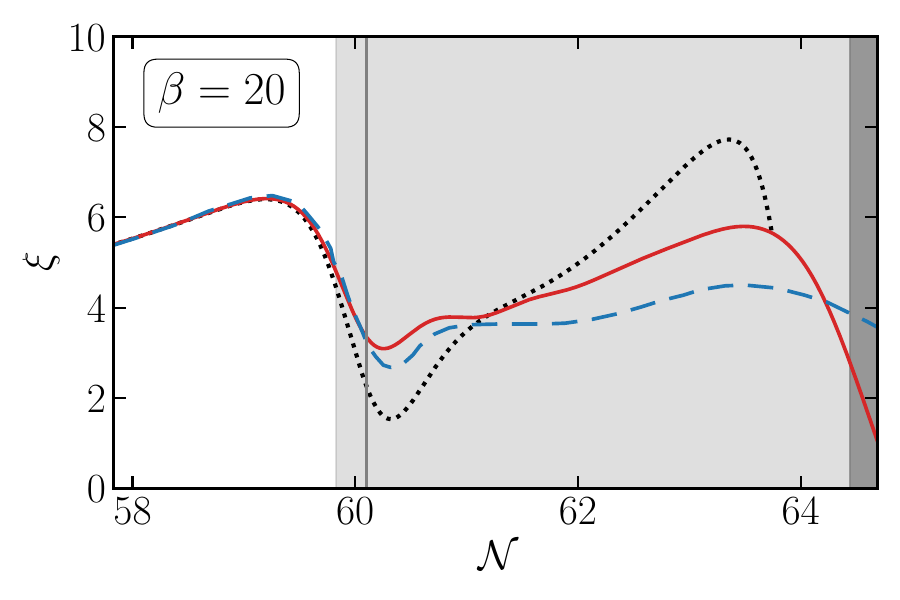}
	\hspace{0.5mm}
	\includegraphics[width=0.32\linewidth]{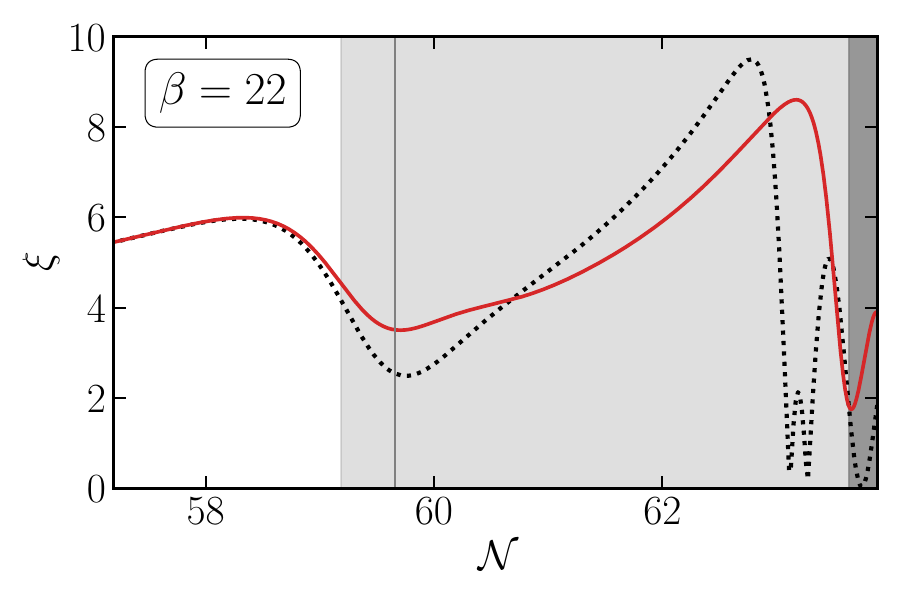}
	\hspace{0.5mm}
	\includegraphics[width=0.32\linewidth]{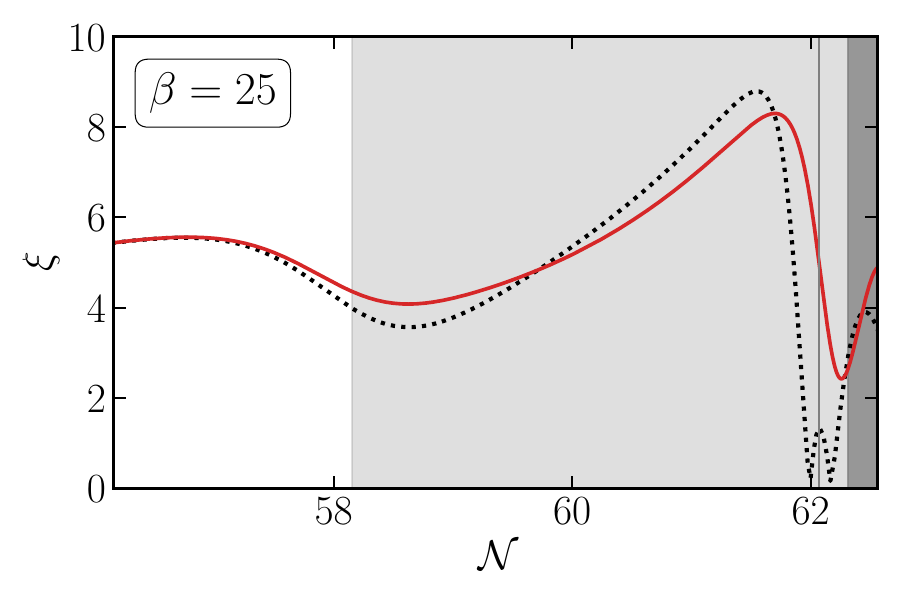}
	\caption{\small 
	Evolution of $\xi$ assuming a homogeneous axion field (dashed black) and perturbatively including axion gradients (red) for different values of $\beta$.
	The light (dark) gray region indicates that the gradient energy of the axion
	exceeds $1\,\%$ ($50\,\%$) of the kinetic energy, while the gray vertical line corresponds to $5\,\%$.
    Wherever possible we compare to the result of the lattice simulation~\cite{Figueroa:2023oxc}.
 }
	\label{fig:xi_pert}
\end{figure}

\begin{figure}[!t]
	\centering
 	\includegraphics[width=0.32\linewidth]{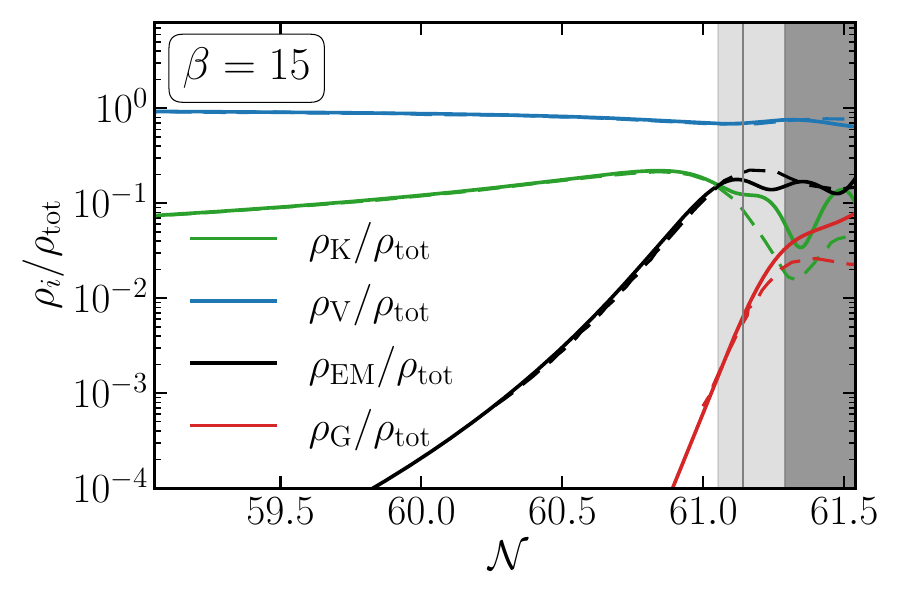}
	\hspace{0.5mm}
 	\includegraphics[width=0.32\linewidth]{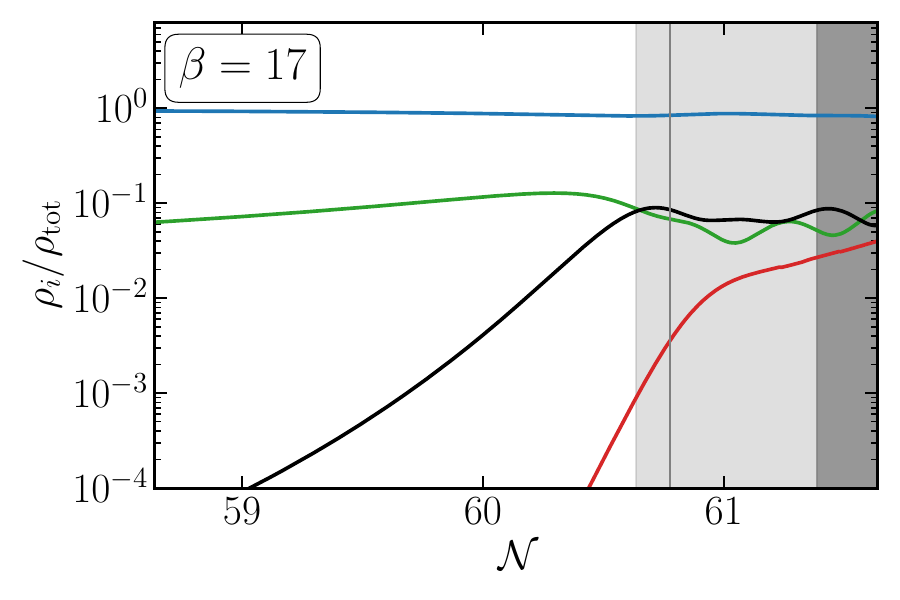}
	\hspace{0.5mm}
 	\includegraphics[width=0.32\linewidth]{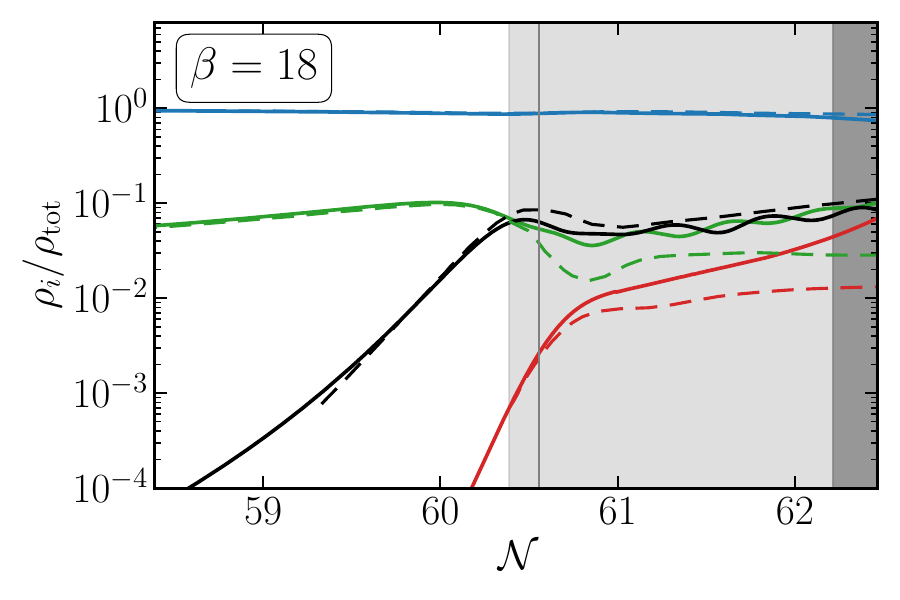}
    \vspace{1mm}
 	\includegraphics[width=0.32\linewidth]{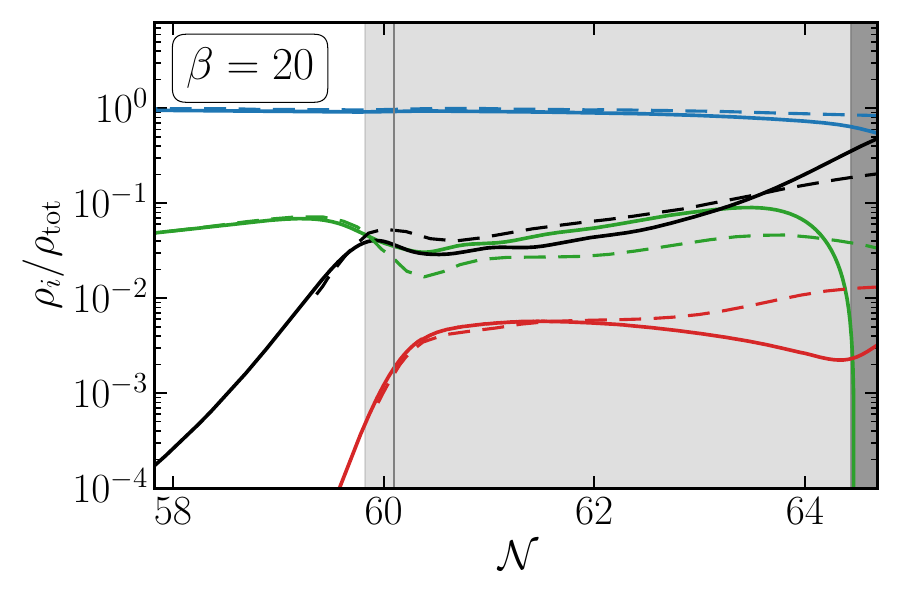}
	\hspace{0.5mm}
 	\includegraphics[width=0.32\linewidth]{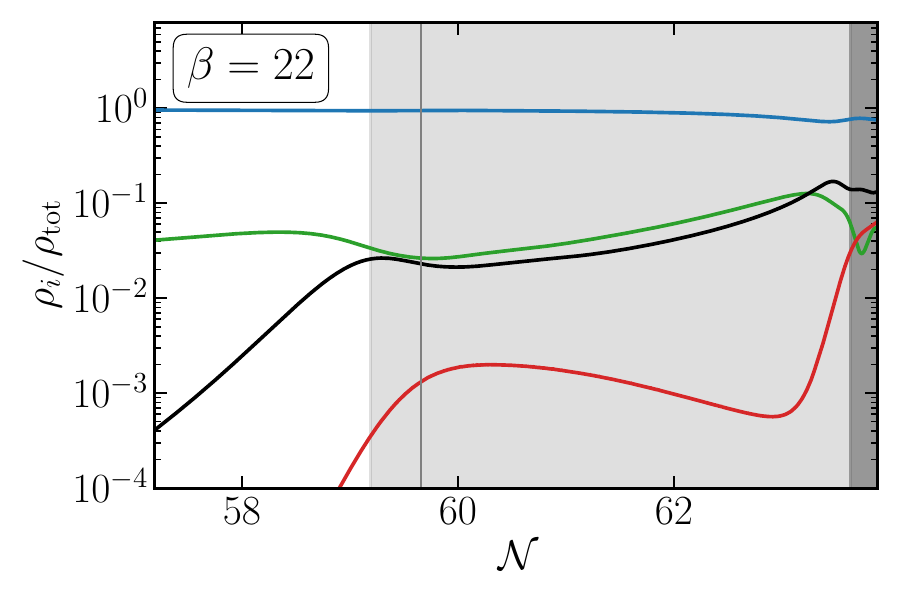}
	\hspace{0.5mm}
 	\includegraphics[width=0.32\linewidth]{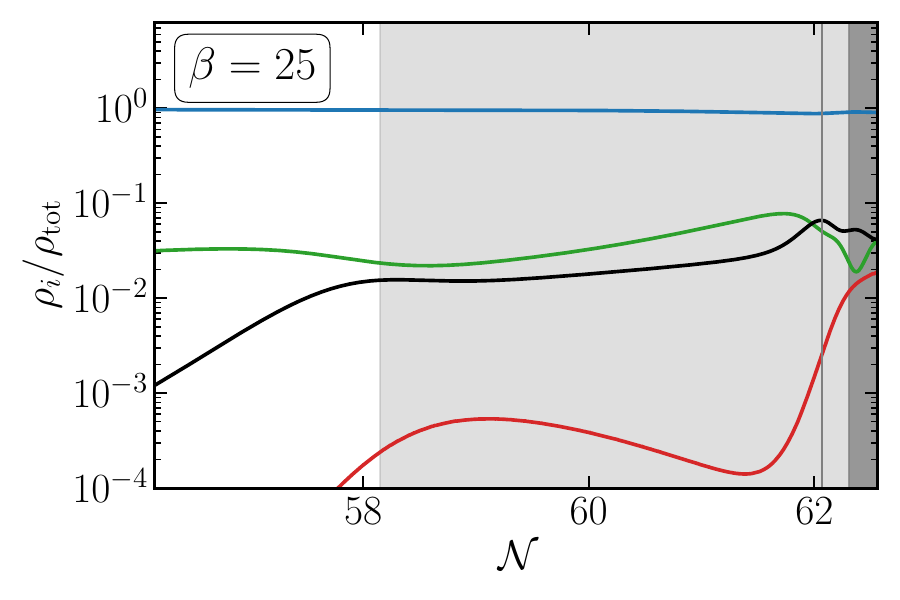}
	\caption{\small 
	Evolution of the energy densities for different values of $\beta$. Gray bands as in Fig.~\ref{fig:xi_pert}.
 Results from lattice simulations for $\beta=15, 18$ and $20$ from Ref.~\cite{Figueroa:2023oxc} are shown in dashed. 
 }
	\label{fig:rho}
\end{figure}

Figs.~\ref{fig:xi_pert} and \ref{fig:rho} show the results of the methodology described in Sec.~\ref{sec:gef} for values of the coupling $\beta$ ranging from 15 to 25. The six panels of Fig.~\ref{fig:xi_pert} focus on the evolution of the parameter $\xi = \beta  \vert\dot \phi \vert/ (2 H M_P)$. In all panels, the dotted black lines show the result for the GEF assuming a homogeneous axion and the solid red lines are new results including perturbatively the axion gradients to first order. Where available, we show for comparison the lattice results obtained in Ref.~\cite{Figueroa:2023oxc} in dashed blue. The gray regions and horizontal lines indicate the quality of the perturbative expansion. In the dark gray region the axion gradient energy exceeds 50\% of the axion kinetic energy, $R_\chi > 0.5$, indicating the non-perturbative regime.
For all panels, we show only 0.25 e-folds of this non-perturbative regime, though we stress that lattice results have shown that inflation can last several e-folds longer. Defining $\Delta\mathcal{N} = \mathcal{N}_{\mathrm{end}} - \mathcal{N}(R_\chi = 0.5)$ leads to $\Delta \mathcal{N} \simeq 1.5$ for $\beta = 15$, $\Delta \mathcal{N} \simeq 3.5$ for $\beta = 18$, and $\Delta \mathcal{N} \simeq 4.5$ for $\beta = 20$~\cite{Figueroa:2023oxc}. 
The light gray region indicates the regime in which the gradient energy is sizeable, i.e. $0.01 < R_\chi < 0.5$, but our perturbative treatment is still valid.
The vertical gray line indicates $R_\chi = 0.05$. We observe that below this value, we recover the full lattice results to good agreement (while the deviation from the homogeneous approximation is already significant). Above this value the leading order correction implemented here is insufficient to fully reproduce the lattice results, but since $R_\chi \ll 1$ a systematic expansion to higher orders in the axion gradients might conceivably achieve this (within the light gray region).

Fig.~\ref{fig:rho} shows the evolution of the kinetic ($\rho_\mathrm{K}$), potential ($\rho_\mathrm{V}$), electromagnetic ($\rho_\mathrm{EM}$) and axion gradient energy ($\rho_\mathrm{G}$) defined as
\begin{align}
	\rho_\mathrm{K} = \frac{1}{2}\left(\dot{\phi}^2 + \mathcal{P}_{\dot{\chi}}^{(0)}\right),
	\quad
	\rho_\mathrm{V} = \frac{m_\phi^2}{2}\left(\phi^2 + \mathcal{P}_{\chi}^{(0)}\right),
	\quad
	\rho_\mathrm{EM} = \frac{1}{2}\left(\mathcal{P}_{E}^{(0)} + \mathcal{P}_{B}^{(0)}\right),
	\quad
	\rho_\mathrm{G} = -\frac{1}{2}\mathcal{P}_{\chi}^{(2)},
\end{align}
for the same values of $\beta$. 
Here we include $\rho_\mathrm{G}$ in the total energy density, 
$\rho_\mathrm{tot} = \rho_\mathrm{K} + \rho_\mathrm{V} + \rho_\mathrm{EM} + \rho_\mathrm{G}$,
although we do not include $\rho_\mathrm{G}$ in the Friedmann equation as we explained above.\footnote{
	Including this term, despite spoiling the consistency of the Friedmann equation, does not affect our result 
	in the regime where we can trust our perturbative treatment of the axion gradient.
}
This figure gives a more detailed view of the importance of the axion gradient energy, showing that in many cases (e.g.\ $\beta = 20,22,25$) there is a prolonged regime in which the axion gradient energy remains at the percent level or below. The opposite is observed for $\beta =15$, where once it becomes relevant, the axion gradient energy rapidly becomes comparable to the other components. The efficiency of the method discussed in Sec.~\ref{sec:gef} allows us to study a wide range of couplings $\beta$, confirming that diverse and complex dynamics occurring in the strong backreaction regime. In particular, possible observable signatures related to the different energy components depend on the coupling $\beta$ in a non-monotonic way.

\begin{figure}[!t]
\centering
\begin{tabular}{cc}
\hspace{-0.5cm} \includegraphics[width=0.49\textwidth]{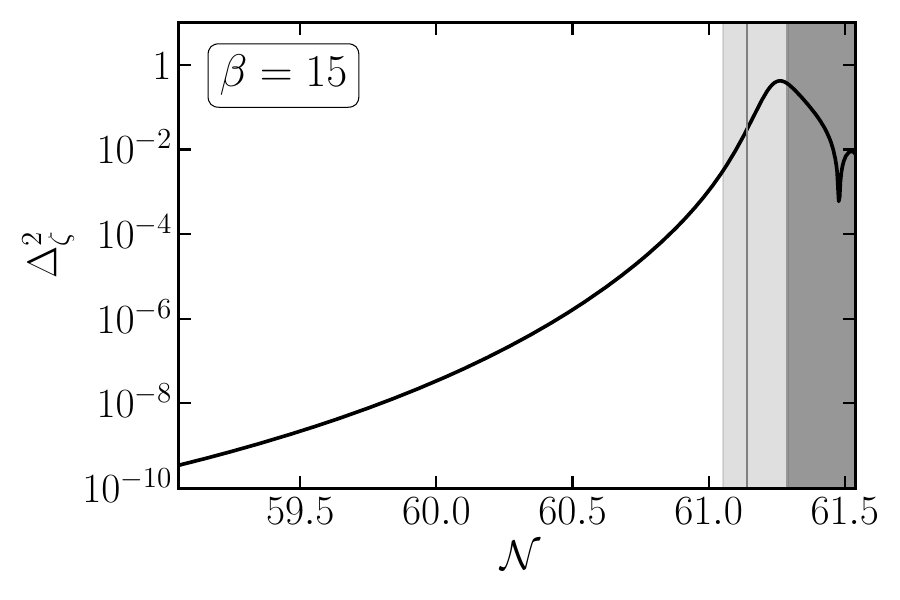} & 
\hspace{-0.5cm} \includegraphics[width=0.49\textwidth]{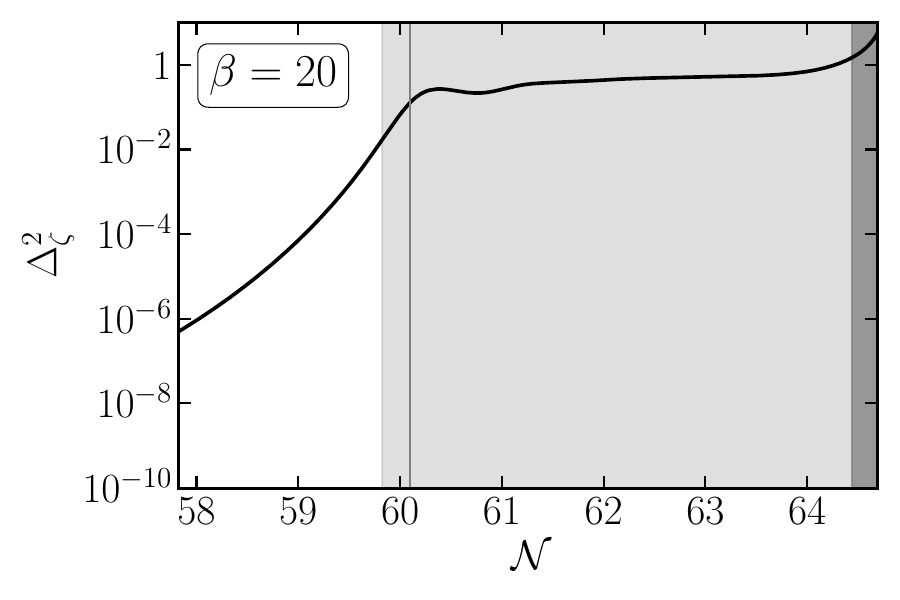}
\end{tabular}
\vspace{-0.4cm}
\caption{\small
    Power spectrum of axion perturbations, $\Delta_\zeta^2 = (H/\dot{\phi})^2 \mathcal{P}_\chi^{(0)}$, for $\beta = 15$ (left) and $\beta = 20$ (right). 
    Large values imply PBH formation, see text for details. 
    Gray bands as in Fig.~\ref{fig:xi_pert}.
}
\label{fig:p_chi}
\end{figure}

As one example of possible observable consequences we show the scalar power spectrum sourced by the axion fluctuations, $\Delta_\zeta^2 = (H/\dot \phi)^2 \mathcal{P}_\chi^{(0)}$ in Fig.~\ref{fig:p_chi}.\footnote{
    Strictly speaking $\mathcal{P}_\chi^{(0)}$ is an integrated quantity and we do not have access to the power spectrum for each different momentum $k$ in our formalism 
    (unless we compute the axion 2-point functions with sufficiently high powers of the spatial derivatives). However, we expect that at each given time the modes with $k \sim k_h$ dominate the integral
    and the time evolution of $\mathcal{P}_X^{(0)}$ roughly corresponds to the spectrum. 
}
Note that this contains only the contribution from the axion fluctuations, and not the contribution from the gauge field energy density fluctuations or metric fluctuations.
We note that even within the perturbative regime of axion gradients we significantly exceed the threshold for primordial black hole (PBH) formation which is estimated to lie around $\Delta_\zeta^2 \sim 10^{-2}$ for a Gaussian distribution and around $\Delta_\zeta^2 \sim 10^{-4}$ for the non-Gaussian spectrum assumed in Ref.~\cite{Linde:2012bt} for axion inflation. PBHs generated in the last few e-folds of inflation are very light and decay rapidly through Hawking radiation leaving no traces in our cosmological history. However, as we see from Fig.~\ref{fig:p_chi}, even for a moderate coupling of $\beta = 15$ the non-perturbative regime of axion gradients is reached before the end of inflation, making a precise prediction of the PBH spectrum impossible with perturbative methods. It is nevertheless interesting to note that even for such small values of $\beta$ (which, in particular, do not produce excessive gravitational waves during preheating~\cite{Adshead:2019lbr,Adshead:2019igv}) our results indicate a phase of PBH formation. Increasing the coupling $\beta$ will typically extend this phase, leading to more massive PBHs with longer lifetimes.
Interestingly, for some parameter choices such as $\beta = 20$, displayed in the right panel, we find indication for scalar perturbations above the PBH formation threshold for a significant range of e-folds within the perturbative regime.

\addtocontents{toc}{\vspace{-0.5em}} 
\section{Discussion and outlook}
\label{sec:conclusion}

The key result of this work is the extension of the gradient expansion formalism (GEF) to perturbatively include axion gradients when evaluating the dynamics of axion inflation. This involves applying the GEF not only to the $2$-point functions but also to higher $p$-point functions, resulting in an expansion in (i) the number of derivatives $n$ acting on the gauge fields in the $2$-point function as in the original GEF, (ii) the order of $p$ of the highest correlation function which is not factorized into lower $p$-point functions and (iii) the number $n_p$ of derivatives in those $p$-point functions. Here we consider the leading order correction due to axion gradients, which requires $p=3$ and $n_3 = n_{\cal B} = 1$. 

This captures several important aspects of the dynamics of the system. For very small values of the expansion parameter $R_\chi$ we recover the results of the homogeneous GEF approximation (i.e. assuming the axion field to be perfectly homogeneous but allowing for gradients in the gauge fields) which in this regime agree with the lattice results capturing the full non-linear dynamics. For $R_\chi \gtrsim 1\%$, the homogeneous GEF results and the lattice results start to diverge, with our perturbative expansion recovering the lattice results. For $R_\chi \gtrsim 5\%$ the leading order corrections included here no longer suffice to accurately track the system, though given the smallness of $R_\chi$ a systematic inclusion of higher order terms might conceivably achieve this. We see no fundamental obstruction to this in this regime. 
In particular, the number of the $3$-point functions scales as $n_\mathcal{B}^3$ 
(or milder by using e.g. the integration by parts) and this eventually limits the inclusion of higher order terms.
However, this happens only when $n_\mathcal{B}$ is sufficiently large and the computational time does not change
drastically by a mild increase of $n_\mathcal{B}$ 
(note that the number of the 2-point functions we numerically follow is $\sim 750$ for
our choice of $n_\mathrm{max} = 250$).
Finally, for $R_\chi \gtrsim 0.5$ perturbativity is violated and a full lattice simulation seems unavoidable. Our method allows to very efficiently identify the regions and provide accurate initial conditions for them, thereby enabling future lattice simulations to focus their computational power on these non-perturbative regimes. 

An accurate calculation of the strong backreaction regime in axion inflation is crucial to exploit its rich phenomenology and conclusively test this inflation model. This includes the remaining duration of inflation once the system enters the strong backreaction regime (see Ref.~\cite{Figueroa:2023oxc}), the magnitude of peaks in the scalar power spectrum as well their position with regards to the end of inflation which are crucial to determine if a sizable amount of primordial black hole are formed and their mass distribution (see Ref.~\cite{Linde:2012bt}) as well as the anisotropic component of the gauge field energy momentum tensor which will determine the gravitational wave spectrum (see Ref.~\cite{Garcia-Bellido:2023ser}). Such results will need to be contrasted with model constraints coming from the gravitational wave production in the preheating era~\cite{Adshead:2019lbr,Adshead:2019igv}. Assuming a simple shape of the scalar potential throughout the inflation and preheating era, these impose stringent bounds on the axion gauge field coupling, which however still allow for an inflationary phase within the strong backreaction regime.  
Currently, only costly lattice simulations can give quantitative answers to all these questions. 

The method proposed here provides a first step towards an efficient way of studying these questions across the parameter space of axion inflation. Several extensions of the scheme proposed here deserve further study, e.g. the inclusion of higher order $p$-point functions and/or derivatives therein as well as including a correction algorithm as proposed in~\cite{vonEckardstein:2023gwk}. We hope that this work will trigger some of these developments.

\paragraph{Acknowledgements}
We thank Kai Schmitz, Oleksandr Sobol and Richard von Eckardstein for fruitful discussions and the very helpful comparisons of our numerical codes, as well as  Dani Figueroa and Ander Urio Garmendia for insightful cross-checks from lattice simulations. We moreover thank Ryo Namba for helpful discussions,
and Kyohei Mukaida for collaboration at the initial stages of this project.
Y.E. and S.S. would like to thank the CERN Theory Department for their hospitality during crucial parts of this project.
The work of Y.E. is supported in part by DOE grant DE-SC0011842.
The work of S.S. received the support of a fellowship from “la Caixa” Foundation (ID 100010434) with fellowship code LCF/BQ/DI19/11730034 and by the Generalitat Valenciana grant PROMETEO/2021/083.

\newpage
\appendix

\addtocontents{toc}{\vspace{0.0em}} 

\section{Equations of motion for $3$-point functions}
\label{app:eom_3pt}

The main goal of this paper is to extend the GEF and include the axion inhomogeneity $\chi(t,\vec{x})$.
Once we include $\chi$, the EOMs become
non-linear in terms of the inhomogeneous quantities, $\chi$, $\vec{E}$ and $\vec{B}$.
As a result, the EOMs of the electromagnetic $2$-point functions no longer form a closed system
and we need to include the evolution of the $3$-point functions.
The EOM of the $3$-point functions then depends on $4$-point functions and
a similar structure persists for higher-point functions.
To truncate this tower of $p$-point functions, we factorize the $4$-point functions as products of the $2$-point functions.
Moreover, for simplicity, we include the $3$-point functions 
with only up to one spatial derivative in the full system.
The latter approximation is expected to be valid when the axion gradient energy is suppressed,
and to check the quality of our approximation, we compute the axion gradient energy by treating it as a perturbation
and monitor its size.

In the following, we list the time evolution equations of the $2$-point and $3$-point functions
that we need in our numerical computation.
These are derived by the repeated use of Eqs.~\eqref{eq:eom_chi}--\eqref{eq:eom_B}.
In this appendix we use the following notation for 3-point functions in the GEF,
\begin{align}
	\mathcal{B}_{f;X}^{(2l; n, m)} &= \frac{1}{a^{n+m + 2l}}
	\vev{ \nabla^{2l} f (\vec{\nabla}\times)^n \vec{X}\cdot 
	(\vec{\nabla}\times )^m \vec{X}}\,,
        \quad
	\mathcal{B}_{f;XY}^{(2l; n, m)} = -\frac{1}{a^{n+m + 2l}}
	\vev{\nabla^{2l} f (\vec{\nabla}\times)^n \vec{X}\cdot 
	(\vec{\nabla}\times)^m \vec{Y}}\,,
\end{align}
with $X, Y = \{E, B\}$ and $f = \{\chi, \dot{\chi}\}$.
These are related to the simplified notation used in the main text (which does not include axion gradients) as
\begin{align}
	\mathcal{B}_{f;E}^{(n)} = \mathcal{B}_{f;E}^{(0; n, 0)}\,,
	\quad
	\mathcal{B}_{f;B}^{(n)} = \mathcal{B}_{f;B}^{(0; n, 0)}\,,
	\quad
	\mathcal{B}_{f;EB}^{(0)} = \mathcal{B}_{f;EB}^{(0; 0, 0)}\,,
	\quad
	\mathcal{B}_{f;EB}^{(1,0)} = \mathcal{B}_{f;EB}^{(0; 1, 0)}\,,
	\quad
	\mathcal{B}_{f;EB}^{(0,1)} = \mathcal{B}_{f;EB}^{(0; 0, 1)}\,.
\end{align}
Notice that
\begin{align}
	\mathcal{B}_{f;E}^{(2l; n, m)} = \mathcal{B}_{f;E}^{(2l; m, n)}\,,
	\quad
	\mathcal{B}_{f;B}^{(2l; n, m)} = \mathcal{B}_{f;B}^{(2l; m, n)}\,,
\end{align}
by definition, but in general
\begin{align}
	\mathcal{B}_{f;EB}^{(2l; n, m)}
	\neq
	\mathcal{B}_{f;EB}^{(2l; m, n)}\,,
\end{align}
 for $n\neq m$.

\subsection{Equations of motion up to one spatial derivative}

As we stated above, we include the $3$-point functions with only up to one spatial derivative.
With this in mind, the electromagnetic $2$-point functions evolve as
\begin{align}
	&\dot{\mathcal{P}}_E^{(n)} + (n+4)H\mathcal{P}_E^{(n)} + 2 \mathcal{P}_{EB}^{(n+1)} 
	- \frac{2\beta}{M_P} 
	\left(\dot{\phi}\mathcal{P}_{EB}^{(n)}  + \mathcal{B}_{\dot{\chi};EB}^{(0; n,0)} \right)
	+ \frac{2\beta}{M_P}\left(\mathcal{B}_{\chi;E}^{(0; n+1,0)} - \mathcal{B}_{\chi;E}^{(0; n,1)}\right)
	= \left[\dot{\mathcal{P}}_E^{(n)}\right]_b\,,
	 \nonumber \\
	&\dot{\mathcal{P}}_B^{(n)} + (n+4)H\mathcal{P}_B^{(n)} - 2 \mathcal{P}_{EB}^{(n+1)}
	= \left[\dot{\mathcal{P}}_B^{(n)}\right]_{b}\,,
	 \\
	&\dot{\mathcal{P}}_{EB}^{(n)} + (n+4)H \mathcal{P}_{EB}^{(n)}
	- \mathcal{P}_E^{(n+1)} + \mathcal{P}_B^{(n+1)} 
	- \frac{\beta}{M_P}\left( \dot{\phi}\mathcal{P}_{B}^{(n)}
	+ \mathcal{B}_{\dot{\chi}; B}^{(0; n,0)}\right)
	- \frac{\beta}{M_P}\left(\mathcal{B}_{\chi;EB}^{(0; 1,n)} - \mathcal{B}_{\chi;EB}^{(0; 0,n+1)}\right)
	= \left[\dot{\mathcal{P}}_{EB}^{(n)}\right]_{b}\,, \nonumber
\end{align}
where it is understood that
\begin{align}
	\mathcal{B}_{f;XY}^{(2l;n,m)} = 0,
	~~\mathrm{for}~~n+m + 2l > 1\,,
\end{align}
in these equations.
The axion $2$-point functions with no spatial derivatives are given by
\begin{align}
	&\dot{\mathcal{P}}_{\chi}^{(0)}
	- 2\mathcal{P}_{\chi\dot{\chi}}^{(0)} = 0\,,
	\nonumber \\
	&\dot{\mathcal{P}}_{\chi\dot{\chi}}^{(0)} + 3H\mathcal{P}_{\chi\dot{\chi}}^{(0)}
	+ m_\phi^2 \mathcal{P}_{\chi}^{(0)}
	+ \frac{\beta}{M_P}\mathcal{B}_{\chi; EB}^{(0; 0,0)} - \mathcal{P}_{\dot{\chi}}^{(0)} = 0\,,
	\nonumber \\
	&\dot{\mathcal{P}}_{\dot{\chi}}^{(0)}
	+ 6H\mathcal{P}_{\dot{\chi}}^{(0)} + 2m_\phi^2 \mathcal{P}_{\chi\dot{\chi}}^{(0)}
	+ \frac{2\beta}{M_P}\mathcal{B}_{\dot{\chi}; EB}^{(0; 0,0)} = 0\,.
\end{align}
The time evolution equations of the $3$-point functions are given as follows.
For $(2l; n, m) = (0; 0, 0)$, we obtain
\begin{align}
	&\dot{\mathcal{B}}_{\chi;E}^{(0; 0,0)} + 4H\mathcal{B}_{\chi;E}^{(0; 0,0)} 
	+ 2\mathcal{B}_{\chi;EB}^{(0; 0,1)}
	- \frac{2\beta}{M_P}\left(\dot{\phi}\mathcal{B}_{\chi;EB}^{(0; 0,0)} 
	+ \mathcal{P}_{\chi\dot{\chi}}^{(0)}\mathcal{P}_{EB}^{(0)}\right)
	- \mathcal{B}_{\dot{\chi};E}^{(0; 0,0)} = 0\,,
	\nonumber \\
	&\dot{\mathcal{B}}_{\chi;B}^{(0; 0, 0)} + 4H\mathcal{B}_{\chi;B}^{(0; 0, 0)} 
	- 2\mathcal{B}_{\chi;EB}^{(0; 1, 0)}
	- \mathcal{B}_{\dot{\chi};B}^{(0; 0, 0)} = 0\,,
	\nonumber \\
	&\dot{\mathcal{B}}_{\chi;EB}^{(0; 0, 0)} + 4H\mathcal{B}_{\chi;EB}^{(0; 0, 0)}
	- \mathcal{B}_{\chi;E}^{(0; 1,0)} + \mathcal{B}_{\chi;B}^{(0;1,0)}
	- \frac{\beta}{M_P}\left(\dot{\phi}\mathcal{B}_{\chi;B}^{(0;0,0)} + \mathcal{P}_{\chi\dot{\chi}}^{(0)}\mathcal{P}_B^{(0)}\right)
	- \mathcal{B}_{\dot{\chi};EB}^{(0; 0, 0)} = 0\,,
\end{align}
and
\begin{align}
	&\dot{\mathcal{B}}_{\dot{\chi};E}^{(0; 0,0)} + 7H\mathcal{B}_{\dot{\chi};E}^{(0; 0,0)} 
	+ m_\phi^2 \mathcal{B}_{\chi;E}^{(0;0,0)}
	+ 2\mathcal{B}_{\dot{\chi};EB}^{(0; 0,1)}
	- \frac{2\beta}{M_P}\left(\dot{\phi}\mathcal{B}_{\dot{\chi};EB}^{(0; 0,0)} 
	+ \mathcal{P}_{\dot{\chi}}^{(0)}\mathcal{P}_{EB}^{(0)}\right)
	+ \frac{2\beta}{3M_P}\mathcal{P}_{EB}^{(0)}\mathcal{P}_E^{(0)} = 0\,,
	\nonumber \\
	&\dot{\mathcal{B}}_{\dot{\chi};B}^{(0; 0, 0)} + 7H\mathcal{B}_{\dot{\chi};B}^{(0; 0, 0)} 
	+ m_\phi^2 \mathcal{B}_{\chi;B}^{(0;0,0)}
	- 2\mathcal{B}_{\dot{\chi};EB}^{(0; 1, 0)}
	+ \frac{2\beta}{3M_P}\mathcal{P}_{EB}^{(0)}\mathcal{P}_B^{(0)} = 0\,,
	\nonumber \\
	&\dot{\mathcal{B}}_{\dot{\chi};EB}^{(0; 0, 0)} + 7H\mathcal{B}_{\dot{\chi};EB}^{(0; 0, 0)}
	+ m_\phi^2 \mathcal{B}_{\chi;EB}^{(0;0,0)}
	- \mathcal{B}_{\dot{\chi};E}^{(0; 1,0)} + \mathcal{B}_{\dot{\chi};B}^{(0;1,0)}
	\nonumber \\
	&~~~~~~~~~~~~~~~~~~~~~~~~~~~~~~~~~~~
	- \frac{\beta}{M_P}\left(\dot{\phi}\mathcal{B}_{\dot{\chi};B}^{(0;0,0)} + \mathcal{P}_{\dot{\chi}}^{(0)}\mathcal{P}_B^{(0)}\right)
	+ \frac{\beta}{3M_P}\left[\left(\mathcal{P}_{EB}^{(0)}\right)^2 + \mathcal{P}_E^{(0)}\mathcal{P}_B^{(0)}\right]
	= 0\,.
\end{align}
Notice, in particular, that the equations contain the products of the electromagnetic $2$-point functions.
This is a result of the factorization of the $4$-point functions.
Assuming the isotropy and Gaussianity of the electromagnetic fields, we can factorize e.g.
\begin{align}
	\vev{(\vec{E}\cdot\vec{B})^2}
	\simeq \frac{4}{3}\left(\mathcal{P}_{EB}^{(0)}\right)^2 + \frac{1}{3}\mathcal{P}_E^{(0)}\mathcal{P}_B^{(0)}\,,
\end{align}
and so on.
For $(2l;n,m) = (0;1,0), (0;0,1)$, we obtain
\begin{align}
	&\dot{\mathcal{B}}_{\chi;E}^{(0;1,0)} + 5H\mathcal{B}_{\chi;E}^{(0;1,0)} 
	- \frac{\beta \dot{\phi}}{M_P}
	\left(\mathcal{B}_{\chi;EB}^{(0;1,0)} + \mathcal{B}_{\chi;EB}^{(0;0,1)}\right)
	- \frac{2\beta}{M_P}\mathcal{P}_{\chi\dot{\chi}}^{(0)}\mathcal{P}_{EB}^{(1)}
	- \mathcal{B}_{\dot{\chi};E}^{(0;1,0)}
	= 0\,,
	\nonumber \\
	&\dot{\mathcal{B}}_{\chi;B}^{(0;1,0)} + 5H\mathcal{B}_{\chi;B}^{(0;1,0)}
	- \mathcal{B}_{\dot{\chi};B}^{(0;1,0)} = 0\,,
	\nonumber \\
	&\dot{\mathcal{B}}_{\chi;EB}^{(0;1,0)} + 5H\mathcal{B}_{\chi;EB}^{(0;1,0)}
	- \frac{\beta}{M_P}
	\left(\dot{\phi}\mathcal{B}_{\chi;B}^{(0;1,0)} + \mathcal{P}_{\chi\dot{\chi}}^{(0)}\mathcal{P}_B^{(1)}\right)
	- \mathcal{B}_{\dot{\chi};EB}^{(0;1,0)}
	= 0\,,
	\nonumber \\
	&\dot{\mathcal{B}}_{\chi;EB}^{(0;0,1)} + 5H\mathcal{B}_{\chi;EB}^{(0;0,1)}
	- \frac{\beta}{M_P}
	\left(\dot{\phi}\mathcal{B}_{\chi;B}^{(0;1,0)} + \mathcal{P}_{\chi\dot{\chi}}^{(0)}\mathcal{P}_B^{(1)}\right)
	- \mathcal{B}_{\dot{\chi};EB}^{(0;0,1)}
	= 0\,,
\end{align}
and
\begin{align}
	&\dot{\mathcal{B}}_{\dot{\chi};E}^{(0;1,0)} 
	+ 8H\mathcal{B}_{\dot{\chi};E}^{(0;1,0)} + m_\phi^2 \mathcal{B}_{\chi;E}^{(0;1,0)}
	\nonumber \\
	&~~~~~~~~~~~~~~~~~~~~~~~~
	- \frac{\beta}{M_P}
	\left(\dot{\phi}
	\left(\mathcal{B}_{\dot{\chi};EB}^{(0;1,0)} + \mathcal{B}_{\dot{\chi};EB}^{(0;0,1)}\right)
	+ 2\mathcal{P}_{\dot{\chi}}^{(0)}\mathcal{P}_{EB}^{(1)}
	\right)
	+ \frac{\beta}{3M_P}\left(\mathcal{P}_{EB}^{(0)}\mathcal{P}_E^{(1)} + \mathcal{P}_E^{(0)}
	\mathcal{P}_{EB}^{(1)}\right)
	= 0\,,
	\nonumber \\
	&\dot{\mathcal{B}}_{\dot{\chi};B}^{(0;1,0)} + 8H\mathcal{B}_{\dot{\chi};B}^{(0;1,0)}
	+ m_\phi^2 \mathcal{B}_{\chi;B}^{(0;1,0)}
	+ \frac{\beta}{3M_P}
	\left(\mathcal{P}_B^{(0)}\mathcal{P}_{EB}^{(1)} + \mathcal{P}_{EB}^{(0)}\mathcal{P}_{B}^{(1)}
	\right)
	= 0\,,
	\nonumber \\
	&\dot{\mathcal{B}}_{\dot{\chi};EB}^{(0;1,0)} + 8H\mathcal{B}_{\dot{\chi};EB}^{(0;1,0)}
	+ m_\phi^2 \mathcal{B}_{\chi;EB}^{(0;1,0)}
	- \frac{\beta}{M_P}\left(\dot{\phi}\mathcal{B}_{\dot{\chi};B}^{(0;1,0)} +
	\mathcal{P}_{\dot{\chi}}^{(0)}\mathcal{P}_B^{(1)}\right)
	+ \frac{\beta}{3M_P}\left(\mathcal{P}_B^{(0)}\mathcal{P}_E^{(1)}
	+ \mathcal{P}_{EB}^{(0)}\mathcal{P}_{EB}^{(1)}\right)
	= 0\,,
	\nonumber \\
	&\dot{\mathcal{B}}_{\dot{\chi};EB}^{(0;0,1)} + 8H\mathcal{B}_{\dot{\chi};EB}^{(0;0,1)}
	+ m_\phi^2 \mathcal{B}_{\chi;EB}^{(0;0,1)}
	- \frac{\beta}{M_P}\left(\dot{\phi}\mathcal{B}_{\dot{\chi};B}^{(0;1,0)} +
	\mathcal{P}_{\dot{\chi}}^{(0)}\mathcal{P}_B^{(1)}\right)
	+ \frac{\beta}{3M_P}\left(\mathcal{P}_E^{(0)}\mathcal{P}_B^{(1)}
	+ \mathcal{P}_{EB}^{(0)}\mathcal{P}_{EB}^{(1)}\right)
	= 0\,.
\end{align}
Finally, the background EOM is given by
\begin{align}
	&0 = \ddot{\phi} + 3H \dot{\phi} + m_\phi^2 \phi
	+\frac{\beta}{M_P}\mathcal{P}_{EB}^{(0)}\,,
	\nonumber \\
	&H^2 = \frac{1}{6M_P^2}\left[\dot{\phi}^2 + \mathcal{P}_{\dot{\chi}}^{(0)}
	+ m_\phi^2\left(\phi^2 + \mathcal{P}_{\chi}^{(0)}\right) + \mathcal{P}_E^{(0)} + \mathcal{P}_B^{(0)}\right]\,,
	\nonumber \\
	&\dot{H} = -\frac{1}{6M_P^2}\left[3\dot{\phi}^2 + 3\mathcal{P}_{\dot{\chi}}^{(0)}
	+ 2\left(\mathcal{P}_E^{(0)} + \mathcal{P}_B^{(0)}\right)\right]\,.
\end{align}
These equations are consistent, even after our approximations, 
in the sense that we can derive the last equation from the others.

\subsection{Equations of motion with two spatial derivatives}

To monitor the size of the axion inhomogeneity, we follow the time evolution of the axion gradient energy,
i.e., $\mathcal{P}_{\chi}^{(2)}$.
For this purpose, we need to follow some of the $3$-point functions with two spatial derivatives.
We treat these quantities as perturbations, by treating the quantities with only up to one spatial derivative
as the source term and ignoring the backreaction of the axion gradient energy.
The relevant equations are given by
\begin{align}
	&\dot{\mathcal{P}}_{\chi}^{(2)} + 2 H \mathcal{P}_{\chi}^{(2)}
	- 2\mathcal{P}_{\chi\dot{\chi}}^{(2)} = 0\,,
	\nonumber \\
	&\dot{\mathcal{P}}_{\chi\dot{\chi}}^{(2)} + 5H\mathcal{P}_{\chi\dot{\chi}}^{(2)}
	+ m_\phi^2 \mathcal{P}_{\chi}^{(2)}
	+ \frac{\beta}{M_P}\mathcal{B}_{\chi; EB}^{(2; 0,0)} - \mathcal{P}_{\dot{\chi}}^{(2)} = 0\,,
	\nonumber \\
	&\dot{\mathcal{P}}_{\dot{\chi}}^{(2)}
	+ 8H\mathcal{P}_{\dot{\chi}}^{(2)} + 2m_\phi^2 \mathcal{P}_{\chi\dot{\chi}}^{(2)}
	+ \frac{2\beta}{M_P}\mathcal{B}_{\dot{\chi}; EB}^{(2; 0,0)} 
	= 0\,.
\end{align}
Obviously we need to compute $\mathcal{B}_{f;EB}^{(2;0,0)}$.
For this, it is enough to consider only the following equations:
\begin{align}
	&\dot{\mathcal{B}}_{\chi;B}^{(2; 0,0)} + 6H\mathcal{B}_{\chi;B}^{(2; 0,0)} 
	- \mathcal{B}_{\dot{\chi};B}^{(2; 0,0)} = 0\,,
	\nonumber \\
	&\dot{\mathcal{B}}_{\chi;EB}^{(2; 0,0)} + 6H\mathcal{B}_{\chi;EB}^{(2; 0,0)}
	- \frac{\beta}{M_P}\left(\dot{\phi} \mathcal{B}_{\chi;B}^{(2; 0,0)} 
	+ \mathcal{P}_{\chi\dot{\chi}}^{(2)}\mathcal{P}_{B}^{(0)}\right)
	- \mathcal{B}_{\dot{\chi};EB}^{(2; 0,0)} 
	= 0\,,
\end{align}
and
\begin{align}
	&\dot{\mathcal{B}}_{\dot{\chi};B}^{(2;0,0)}
	+ 9H\mathcal{B}_{\dot{\chi};B}^{(2;0,0)}
	+ m_\phi^2 \mathcal{B}_{\chi;B}^{(2;0,0)}
	- \frac{2\beta}{3M_P}\left(\mathcal{P}_{EB}^{(0)}\mathcal{P}_B^{(2)}
	+ \mathcal{P}_B^{(0)}\mathcal{P}_{EB}^{(2)} - \mathcal{P}_{EB}^{(1)}\mathcal{P}_B^{(1)}
	\right)
	= 0\,,
	\nonumber \\
	&\dot{\mathcal{B}}_{\dot{\chi};EB}^{(2;0,0)}
	+ 9H\mathcal{B}_{\dot{\chi};EB}^{(2;0,0)}
	+ m_\phi^2 \mathcal{B}_{\chi;EB}^{(2;0,0)}
	- \frac{\beta}{M_P}\left(\dot{\phi}\mathcal{B}_{\dot{\chi};B}^{(2;0,0)} 
	+ \mathcal{P}_{\dot{\chi}}^{(2)}\mathcal{P}_B^{(0)}\right)
	 \\
	&~~~~~
	- \frac{\beta}{3M_P}
	\left[\mathcal{P}_B^{(0)}\mathcal{P}_{E}^{(2)}
	+ \mathcal{P}_E^{(0)}\mathcal{P}_B^{(2)}
	+ 2\mathcal{P}_{EB}^{(0)}\mathcal{P}_{EB}^{(2)}
	-\mathcal{P}_B^{(1)}\mathcal{P}_E^{(1)}
	-\left(\mathcal{P}_{EB}^{(1)}\right)^2
	-\frac{\beta^2}{3M_P^2} \mathcal{P}_{\chi}^{(2)}
	\left(\mathcal{P}_B^{(0)}\right)^2
	\right]
	= 0\,, \nonumber
\end{align}
where we used the factorization
\begin{align}
	\vev{\partial_i (\vec{E}\cdot\vec{B}) \partial_i (\vec{X}\cdot\vec{Y})}
	&\simeq
	\frac{1}{3}\left(\vev{\vec{E}^{(1)}\cdot \vec{X}^{(1)}} 
	+ \vev{(\vec{\nabla}\cdot\vec{E})(\vec{\nabla}\cdot\vec{X})}\right)
	\vev{\vec{B}\cdot\vec{Y}}
	- \frac{1}{6}\vev{\vec{E}^{(1)}\cdot\vec{Y}}\vev{\vec{B}\cdot\vec{X}^{(1)}}
	\nonumber \\
	&
	+ (X\leftrightarrow Y) + (E\leftrightarrow B) + (X\leftrightarrow Y, E\leftrightarrow B)\,.
\end{align}
Here we use the short-hand notation $\vec{X}^{(n)} = (\vec{\nabla}\times )^n \vec{X}$.
These are closed by themselves, and hence we do not need to compute e.g. 
$\mathcal{B}_{f;E}^{(2;0,0)}$ nor $\mathcal{B}_{f;EB}^{(0;2,0)}$.

\subsection{Numerical implementation}

The system of the equations described above has to be solved numerically. 
It is then convenient to apply the following variable re-definition 
\begin{align}
\begin{split}
    &\bar{\phi} = \frac{\phi}{M_p}\,,\,\, 
    \bar{\mathcal{P}}_X^{(n)} = \frac{\mathcal{P}_X^{(n)}}{H_0^4 \left(k_h/a\right)^n}\,,\,\, 
    \bar{t} = H_0 t\,,\,\, \bar{H} = \frac{H}{H_0}\,,\,\, \bar{m} = \frac{m}{H_0}\,,\,\, 
    \frac{\bar{k}_h}{\bar{a}} = \frac{k_h}{a H_0}\,,\,\, \\
    &\bar{\mathcal{P}}_\chi^{(n)} = \frac{\mathcal{P}_\chi^{(n)}}{M_p^2 \left( k_h/a \right)^n}\,,\,\,
    \bar{\mathcal{P}}_{\dot{\chi}}^{(n)} = \frac{\mathcal{P}_{\dot{\chi}}^{(n)}}{M_p^2 H_0^2 \left( k_h/a \right)^n}\,, \,\,
    \bar{\mathcal{P}}_{\chi\dot{\chi}}^{(n)} = \frac{\mathcal{P}_{\chi\dot{\chi}}^{(n)}}{M_p^2 H_0 \left( k_h/a \right)^n}\,,\,\, \\
    &\bar{\mathcal{B}}_{\chi; X}^{(2l;n,m)} = \frac{\mathcal{B}_{\chi;X}^{(2l;n,m)}}{M_p H_0^4 \left( k_h/a \right)^{2l+n+m}}\,,\,\,
    \bar{\mathcal{B}}_{\dot{\chi};X}^{(2l;n,m)} = 
    \frac{\mathcal{B}_{\dot{\chi};X}^{(2l;n,m)}}{M_p H_0^5 \left( k_h/a \right)^{2l+n+m}}\,,
\end{split}
\end{align}
where $H_0$ is the Hubble parameter at the beginning of the simulation.
The normalization of the power spectra and bispectra by respective powers of $(k_h/a)$ is found to be crucial to numerically evolve the tower of the $2$-point functions of Eqs.~\eqref{eq:PEn}--\eqref{eq:PEBn} to high $n$.
Indeed, we empirically know that $(k_h/aH_0) \sim \mathcal{O}(0.01)$ at the end of inflation,
and hence a different normalization (e.g. by $H_0$ instead of $k_h/a$) can cause an extremely tiny numerical value
for high enough $n$.
Our normalization ensures that each term of the gradient expansion of a given $p$-point function is of the same order, resulting in a more numerically stable system.

We use the multistep Adams method of the GNU scientific library implemented in C\texttt{++} which integrates the full system in $\mathcal{O}(1\,\mathrm{s})$.
For all plots in this paper we take the initial conditions and parameters following Ref.~\cite{Gorbar:2021rlt}, i.e.,
\begin{align}
	\phi = -15.55\,M_P\,,
	\quad
	\dot{\phi} = \sqrt{\frac{2}{3}}m_\phi M_P\,,
	\quad
	a = 1\,,
	\quad
	\mathcal{P}_X^{(n)} = 0\,,
	\quad
	\mathcal{B}_{f;X}^{(2l;n,m)} = 0\,,
\end{align}
at the beginning of our simulation, with the reduced Planck mass $M_p = 2.435 \times 10^{18}\,\mathrm{GeV}$.
We take the axion mass as
\begin{align}
	\frac{m_\phi}{M_P} = 6.16\times 10^{-6}\,,
\end{align}
following Ref.~\cite{Figueroa:2023oxc}.
The initial Hubble parameter $H_0$ is computed as
\begin{align}
	H_0 = \sqrt{\frac{\dot{\phi}^2 + m_\phi^2 \phi^2}{6M_P^2}}\,.
\end{align}
Throughout this paper, we follow the electromagnetic 2-point function up to $n = n_\mathrm{max} = 250$,
and express those with $n = n_\mathrm{max} + 1$ by the truncation relation.

\section{Truncation relation}
\label{app:trunc}

\begin{figure}[!t]
\centering
\begin{tabular}{cc}
\hspace{-0.5cm} \includegraphics[width=0.49\textwidth]{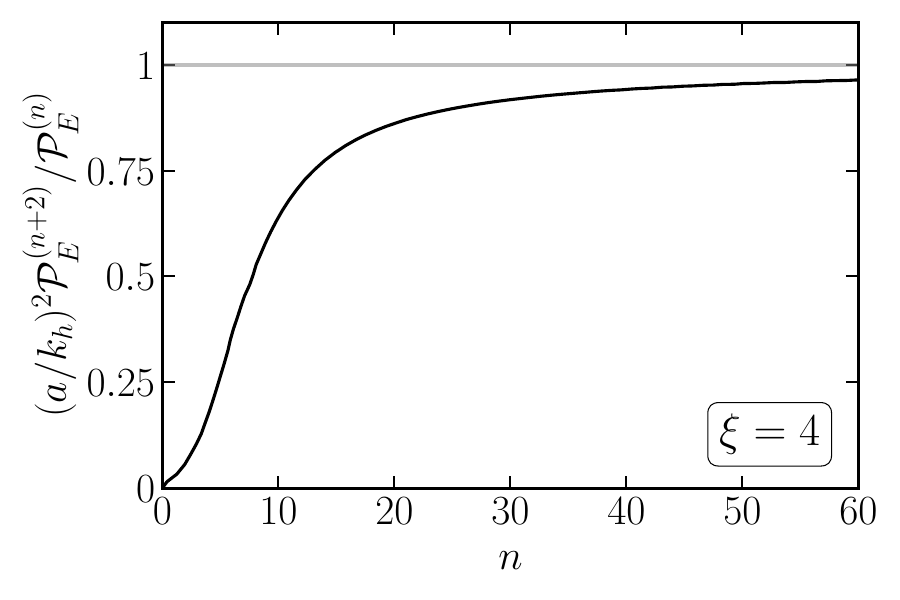} & 
\hspace{-0.5cm} \includegraphics[width=0.49\textwidth]{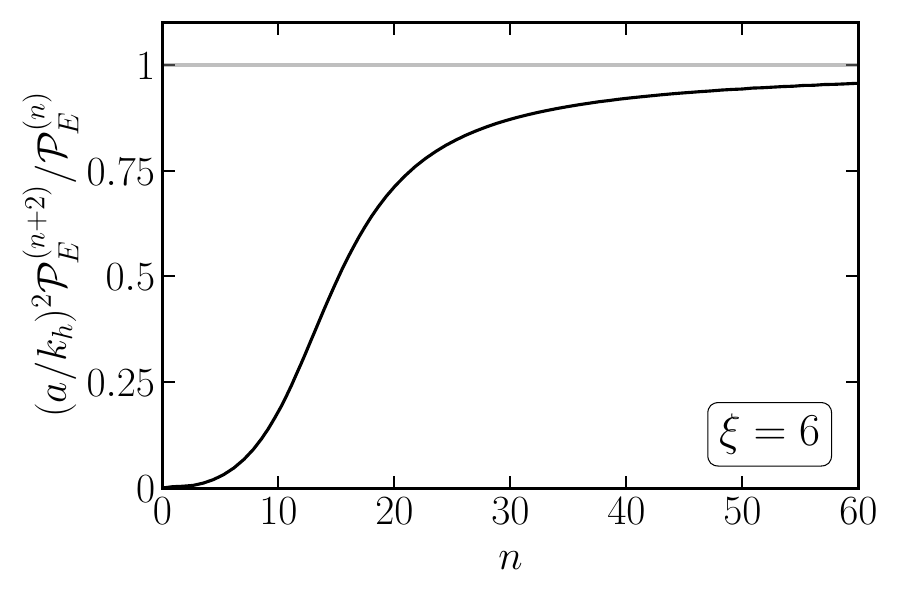}
\end{tabular}
\vspace{-0.4cm}
\caption{\small
For constant values of $\xi$ the truncation relation~\eqref{eq:trunc} holds asymptotically for large $n$, see Eq.~\eqref{app:eq:trunc0}.
The left (right) panel depicts the exemplary case of $\xi = 4$ ($\xi = 6$), the horizontal line indicates the truncation relation assumed in the GEF in \cite{Gorbar:2021rlt}.
}
\label{app:fig:trunc0}
\end{figure}

One of the key approximations in the gradient expansion formalism is the truncation of the tower of equations~\eqref{eq:eom_PEn}-\eqref{eq:eom_PEBn} through the relation~\eqref{eq:trunc}. In this appendix we demonstrate that this relation holds asymptotically for large $n$ and approximately constant $\xi$. In contrary to the derivation given in Ref.~\cite{Gorbar:2021rlt}, this does not rely on approximating the gauge field spectra by a power law around $k_h$, and explains why values of $n = {\cal O}(100)$ are required to obtain accurate results in the GEF. It also illustrates how when $\xi$ changes rapidly, imposing this truncation relation introduces sizable errors, as observed also in Ref.~\cite{vonEckardstein:2023gwk}. Finally, we will introduce an improved truncation relation which mitigates some of these effects. 
For simplicity, the discussion in this appendix is focused on the application of the GEF assuming a homogeneous axion field, as in Ref.~\cite{Gorbar:2021rlt}. The arguments can immediately be extended to include axion gradients as described in Sec.~\ref{sec:gef}.

Let us begin by studying the truncation relation assuming constant $\xi$.
In the slow-roll approximation with constant $\xi$ the gauge field $A_k(\tau)$ mode function can be determined analytically by solving
\begin{align}
    \left(\partial_x^2 + 1 \pm \frac{2\xi }{x} \right) \omega_{\mp} = 0\,,
\end{align}
where we defined $\omega_\pm = \sqrt{2k} A_k^\pm(\tau)$ and $x = - k \tau$, see Eq.~\eqref{eq:A_mode_ODE}.
Matching to the Bunch-Davies vacuum fixes
\begin{align}
A_k^+(\tau) = \frac{1}{\sqrt{2k}} e^{\pi \xi/2} W_{-i \xi, 1/2}(2 i k \tau)\,.
\end{align}
The mode $A_k^-$ does not experience a tachyonic enhancement and can thus be safely neglected.
A reasonable approximation of the mode function $A_k^+$ in the IR regime ($x\to 0$), which dominates the contribution to the $n$-point functions,  is given by
\begin{align}
A_k^+(\tau) \sim \frac{1}{\sqrt{2k}} \frac{e^{\pi \xi/2}}{\Gamma(1 + i \xi)} e^{-\xi \sqrt{-k\tau}}\,.
\end{align}
For the $n$-point function of the $E$-field, see Eq.~\eqref{eq:PEn_mode_integral}, the relevant quantity is
\begin{align}
\left| \frac{\mathrm{d} A_k^+(\tau)}{\mathrm{d}\tau} \right| = \left| \frac{1}{\Gamma(1 + i \xi)} \right| \frac{1}{2\sqrt{2}} \frac{\xi e^{\pi \xi/2}}{\sqrt{-\tau}} e^{-\xi\sqrt{-k \tau} }\,.
\end{align}
The integral of Eq.~\eqref{eq:PEn_mode_integral} can then be done analytically and evaluates to
\begin{align}
\mathcal{P}_E^{(n)} = \frac{4^{-4-n} H^{4+n}}{2 \pi^2 \vert \Gamma(1+i\xi)\vert^2} e^{\pi \xi} \xi^{-2 (2+n)} \left( \Gamma(6+2n) - \Gamma(6+2n, 2\xi \sqrt{-k\tau}) \right)\,.
\end{align}
The ratio of $\mathcal{P}_E^{(n+2)}/\mathcal{P}_E^{(n)}$ is thus
\begin{align}
\label{app:eq:trunc0}
\frac{\mathcal{P}_E^{(n+2)}}{\mathcal{P}_E^{(n)}}  \frac{a^2}{k_h^2}  = \frac{\bar{\mathcal{P}}_E^{(n+2)}}{\bar{\mathcal{P}}_E^{(n)}}   = \frac{1}{(2 \xi)^6} \frac{\Gamma( 10 + 2n ) - \Gamma( 10+2n,2\sqrt{2} \xi^{3/2} )}{\Gamma(6 + 2n) - \Gamma(6+2n, 2\sqrt{2} \xi^{3/2})} \,,
\end{align}
where as in the main text, we have introduced the notation $\bar{\mathcal{P}}_X^{(n)} \equiv \mathcal{P}_X^{(n)}/H_0^4 (k_h/a)^n$.
For large $n$, and any $\xi$, the ratio of Eq.~\eqref{app:eq:trunc0} asymptotically reaches unity, 
\begin{align}
\mathrm{lim}_{n \to \infty} \frac{\mathcal{\bar{P}}_E^{(n+2)}}{\mathcal{\bar{P}}_E^{(n)}}  = 1\,,
\end{align}
with a deviation of $5\,\%$ for $n = 45$ ($n = 55$) for $\xi = 4$ ($\xi = 6$),  see Fig.~\ref{app:fig:trunc0}. 
This, on the one hand confirms the truncation relation chosen in~\cite{Gorbar:2021rlt} in the constant $\xi$ limit, while on the other hand explains why large values of $n$ are required in order for this truncation relation to become relatively accurate. 
Similar conclusion can also be drawn for $\mathcal{\bar{P}}_B^{(n)}$ and $\mathcal{\bar{P}}_{EB}^{(n)}$.

\begin{figure}[t]
\centering
\begin{tabular}{cc}
\hspace{-0.5cm} \includegraphics[width=0.49\textwidth]{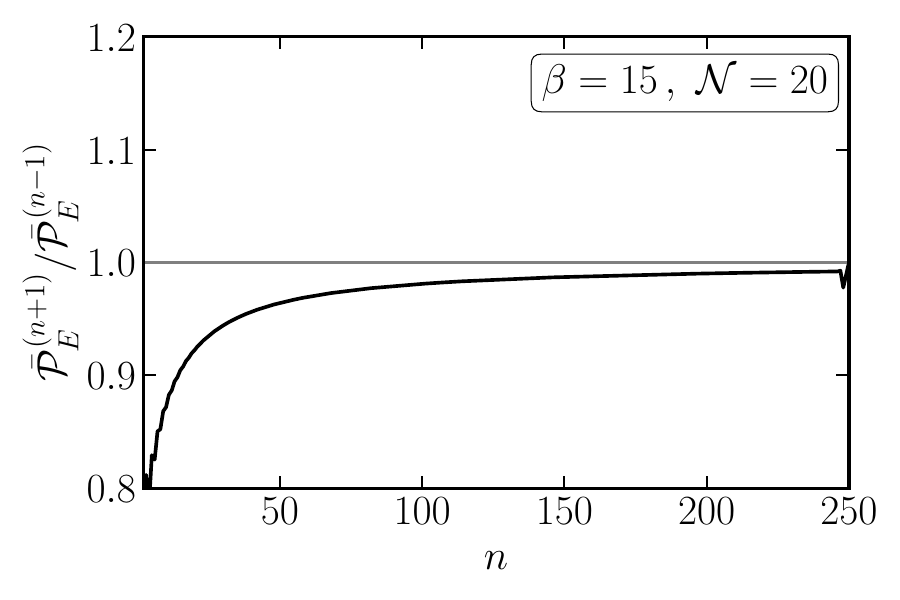} & 
\hspace{-0.5cm} \includegraphics[width=0.49\textwidth]{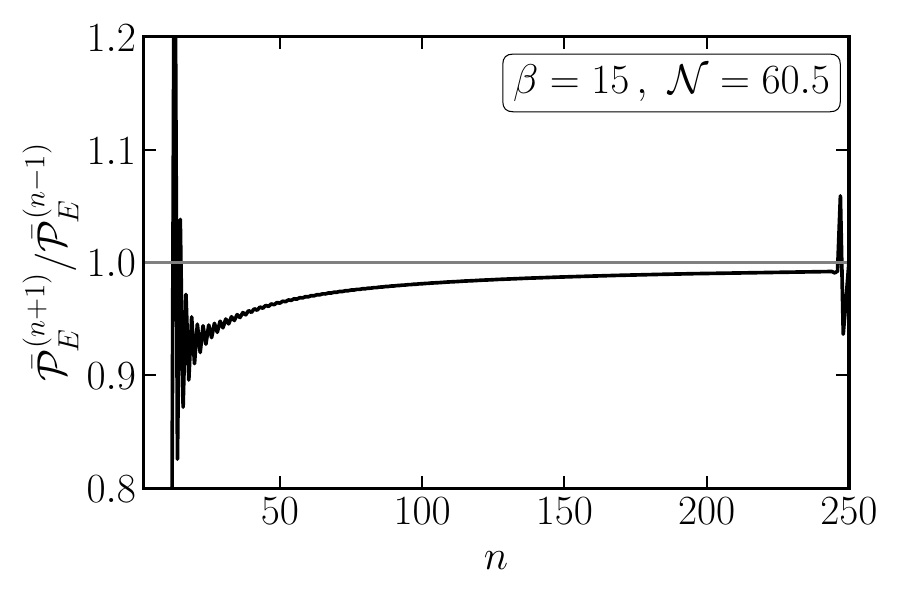}
\end{tabular}
\vspace{-0.4cm}
\caption{\small
Truncation relation~\eqref{eq:trunc} evaluated numerically for fixed $\beta = 15$, showing good agreement for $n \gtrsim 100$ when the backreaction is weak (left panel) and mild (right panel).
The horizontal line indicates the truncation relation assumed in the GEF.
}
\label{app:fig:trunc_numeric_fixed_N}
\end{figure}

The truncation relation remains a good approximation for slowly varying $\xi$, as can be seen in Fig.~\ref{app:fig:trunc_numeric_fixed_N}
which displays the result of a fully numerical solution of the GEF.
In particular, we show the ratio of $\mathcal{\bar{P}}_E^{(n+1)}/\mathcal{\bar{P}}_E^{(n-1)}$ for fixed e-fold $\mathcal{N}$ and $\beta = 15$.
The left panel shows an example from the weak backreaction regime, where $\xi$ changes only slowly and hence the truncation relation is relatively accurate at large $n$. On the other hand, the right panel shows the situation when the inflaton field velocity begins to undergo a more rapid change.
In this case, we observe that the truncation relation remains a good approximation for a large range of $n \gtrsim 50$, although at $n \sim n_\mathrm{max} = 250$ the onset of a violation of this relation can be observed.

\begin{figure}[t]
\centering
\begin{tabular}{cc}
\hspace{-0.5cm} \includegraphics[width=0.49\textwidth]{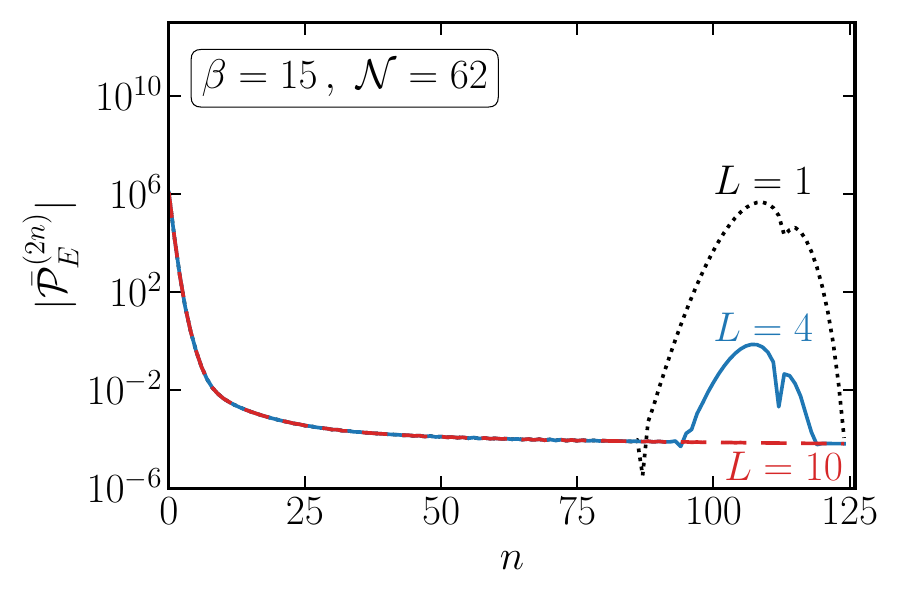} & 
\hspace{-0.5cm} \includegraphics[width=0.49\textwidth]{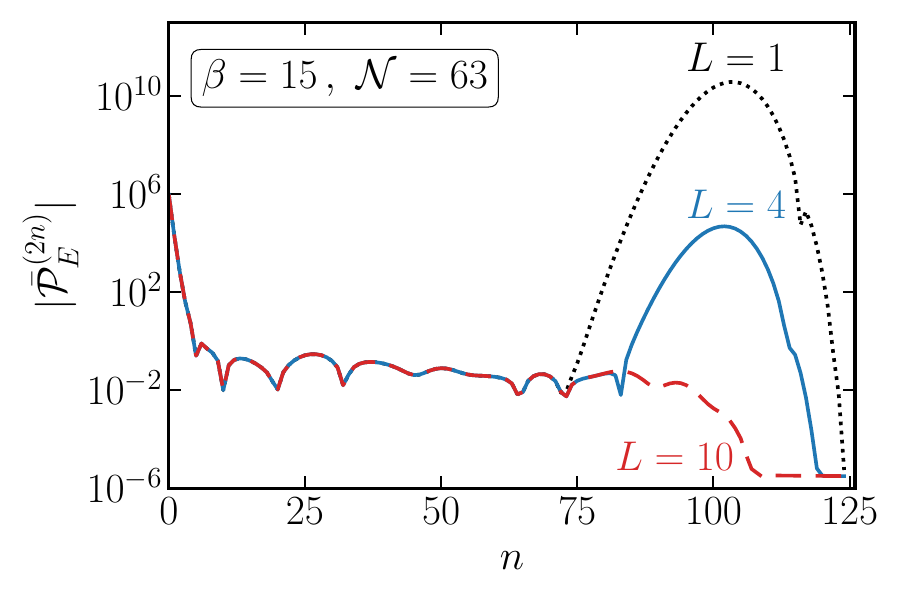}
\end{tabular}
\vspace{-0.4cm}
\caption{\small
GEF tower for rapidly changing $\xi$ using the truncation relation~\eqref{eq:trunc} (dotted black) and the improved truncation relation~\eqref{eq:trunc_improved} with involving the highest four (solid blue) and ten (dashed red) even powers of the GEF tower.
}
\label{app:fig:trunc_numeric_fixed_N_strong_backreaction}
\end{figure}

This violation becomes more dramatic when inflaton velocity changes, and in particular drops, rapidly. This is displayed by the dotted black curve in Fig.~\ref{app:fig:trunc_numeric_fixed_N_strong_backreaction} which shows ${\cal \bar{P}}_E^{(2n)}$ at a slightly later point in time, ${\cal N} = 62$ for different values of $n$. For $50 \lesssim 2n \lesssim 150$ we observe a plateau in ${\cal \bar{P}}_E^{(2n)}$, indicating that the truncation relation~\eqref{eq:trunc} is well satisfied. However, at larger values of $n$ we see very large deviations from the truncation relation, propagating over time down to lower values of $n$. If left unchecked, this will finally affect the $n=0$ components of the GEF tower. Similar effects were recently observed in Ref.~\cite{vonEckardstein:2023gwk}, and similarly to those results we find that these difficulties occur when $\xi$ drops rapidly and $k_h$ features a plateau, see Eq.~\eqref{eq:def_kh}, and the boundary terms are zero. 

To counter this issue, we propose the improved truncation relation~\eqref{eq:trunc_improved}. Averaging over several $\mathcal{\bar{P}}_E^{(n + 1 - 2l)}$ to determine the truncation relation for $\mathcal{\bar{P}}_E^{(n+1)}$,
\begin{align}
    \bar{\mathcal{P}}_X^{(n_\mathrm{max}+1)}
	= \sum_{l = 1}^L (-1)^{l - 1} \begin{pmatrix}  L \\ l \end{pmatrix}
	 \bar{\mathcal{P}}_X^{(n_\mathrm{max}+1-2l)}  \nonumber
\end{align}
gives a more robust procedure, as shown by the colored curves in Fig.~\ref{app:fig:trunc_numeric_fixed_N_strong_backreaction}. 
Compared to the original truncation relation (dotted black), we see that the striking and unphysical feature (`bump') at large $n$ has largely disappeared, at no additional computational cost.

\begin{figure}[t]
\centering
\begin{tabular}{cc}
\hspace{-0.5cm} \includegraphics[width=0.49\textwidth]{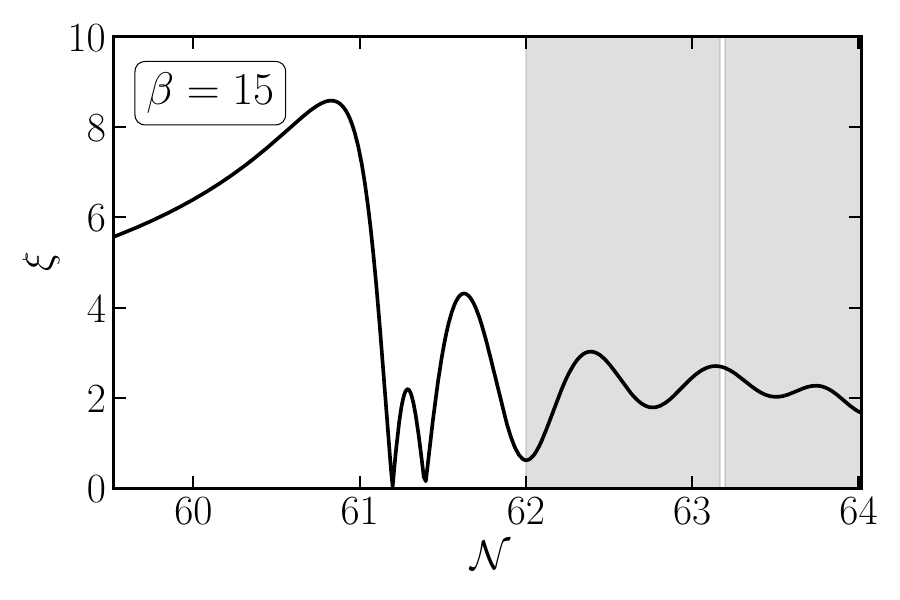} & 
\hspace{-0.5cm} \includegraphics[width=0.49\textwidth]{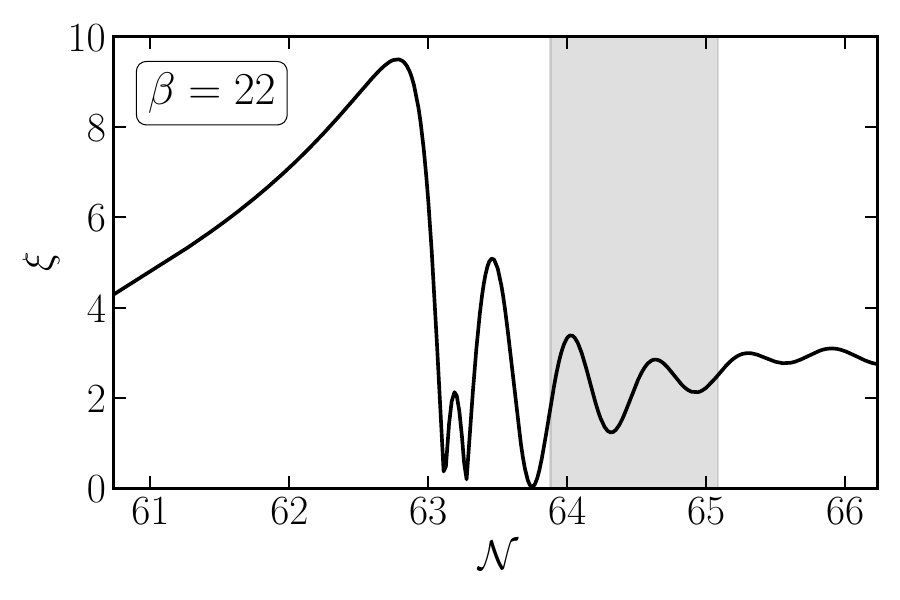}
\end{tabular}
\vspace{-0.4cm}
\caption{
Evolution of $\xi$ for $\beta = 15$ (left) and $\beta = 22$ (right), assuming a homogeneous axion field. 
The gray regions indicate a violation of the truncation relation above $10\%$, see text for details.
}
\label{app:fig:xi_truncation_breakdown}
\end{figure}

Nevertheless, smaller unphysical features (`wiggles') remain, which over time can numerically destabilize the system (see right panel).
In particular, $\mathcal{P}_E^{(2n)}$ should always be positive, while the `wiggles' contain negative valued $\mathcal{P}_E^{(2n)}$.
Similar observations hold for the other $2$-point functions $\mathcal{\bar{P}}^{(n)}_B$ and $\mathcal{\bar{P}}^{(n)}_{EB}$ as well as for odd powers of gradients (although in the latter two cases positivity is not necessarily required).
We were not able to conclusively determine the origin of these `wiggles' nor find a strategy to remove them. 
To illustrate their relevance, the grey shaded regions Fig.~\ref{app:fig:xi_truncation_breakdown} indicate the regions in which (even with the use of Eq.~\eqref{eq:trunc_improved} with $L=10$) we obtain $| \bar{\mathcal{P}}_X^{(2n + 2)}/\bar{\mathcal{P}}_X^{(2n)} - 1| > 0.1$, i.e. a significant violation of the truncation relation, for some values of $50 \leq n \leq 75$ with $X = \{E, B \}$. As can be seen, this typically happens after a sharp drop in $\xi$. For a milder evolution of $\xi$, e.g. for $\beta = 18$ or $20$, this issue does not arise.
In this context, it may be interesting to further study the proposal given in Ref.~\cite{vonEckardstein:2023gwk} based on re-initializing the GEF through the mode-by-mode method (which in turn only requires the input of the $n=0$ mode). 
Since it takes some time for these unphysical effects to propagate to the $n=0$ mode, this procedure might further improve the situation. 
To our understanding, this method was initially proposed to remove the `bump' feature, which Eq.~\eqref{eq:trunc_improved} achieves more efficiently. However, a similar method may prove useful to address the remaining issue of the `wiggles'.

So far, the discussion in this appendix has focused on the homogeneous axion case. Despite the difficulties mentioned above, in practice, for the quadratic scalar potential studied here, the numerical issues in the higher orders of the GEF tower do not propagate to the lowest orders for any coupling $\beta$ considered here before the end of inflation (see however~\cite{vonEckardstein:2023gwk} for different background dynamics). Moreover, including the axion gradients, the remaining difficulties with maintaining a stable truncation relation only occur in the non-perturbative regime, i.e. for $R_\chi > 0.5$, for all values of $\beta$ considered, and do hence not seem to pose a problem within the region of validity of the perturbative method proposed here.
In this sense, the methods employed here ensure sufficient stability of the GEF scheme to study the perturbative regime of axion gradients which is the main goal of this work.

In summary, this appendix clarifies the origin of the truncation relation of the GEF as the asymptotic limit of the GEF tower for large $n$ and approximately constant $\xi$. This sheds some light on the limitations of the GEF formalism for rapidly varying $\xi$, and prompts the introduction of an improved truncation relation. This increases the stability and range of validity of the GEF approach. We point out a remaining instability which however in practice does not impact the results of this work as it does not occur within the perturbative regime of axion inhomogeneities.

%
%

\small
\bibliographystyle{utphys}
\bibliography{biblio}

\providecommand{\href}[2]{#2}\begingroup\raggedright\begin{thebibliography}{10}

\bibitem{Planck:2018jri}
{\bfseries Planck} Collaboration, Y.~Akrami {\em et~al.}, ``{Planck 2018
  results. X. Constraints on inflation},''
  \href{http://dx.doi.org/10.1051/0004-6361/201833887}{{\em Astron. Astrophys.}
  {\bfseries 641} (2020) A10},
  \href{http://arxiv.org/abs/1807.06211}{{\ttfamily arXiv:1807.06211
  [astro-ph.CO]}}.

\bibitem{Planck:2018vyg}
{\bfseries Planck} Collaboration, N.~Aghanim {\em et~al.}, ``{Planck 2018
  results. VI. Cosmological parameters},''
  \href{http://dx.doi.org/10.1051/0004-6361/201833910}{{\em Astron. Astrophys.}
  {\bfseries 641} (2020) A6}, \href{http://arxiv.org/abs/1807.06209}{{\ttfamily
  arXiv:1807.06209 [astro-ph.CO]}}. [Erratum: Astron.Astrophys. 652, C4
  (2021)].

\bibitem{Freese:1990rb}
K.~Freese, J.~A. Frieman, and A.~V. Olinto, ``{Natural inflation with pseudo -
  Nambu-Goldstone bosons},''
\href{http://dx.doi.org/10.1103/PhysRevLett.65.3233}{{\em Phys. Rev. Lett.}
  {\bfseries 65} (1990) 3233--3236}.

\bibitem{Anber:2009ua}
M.~M. Anber and L.~Sorbo, ``{Naturally inflating on steep potentials through
  electromagnetic dissipation},''
  \href{http://dx.doi.org/10.1103/PhysRevD.81.043534}{{\em Phys. Rev. D}
  {\bfseries 81} (2010) 043534},
  \href{http://arxiv.org/abs/0908.4089}{{\ttfamily arXiv:0908.4089 [hep-th]}}.

\bibitem{Linde:2012bt}
A.~Linde, S.~Mooij, and E.~Pajer, ``{Gauge field production in supergravity
  inflation: Local non-Gaussianity and primordial black holes},''
  \href{http://dx.doi.org/10.1103/PhysRevD.87.103506}{{\em Phys. Rev. D}
  {\bfseries 87} no.~10, (2013) 103506},
  \href{http://arxiv.org/abs/1212.1693}{{\ttfamily arXiv:1212.1693 [hep-th]}}.

\bibitem{Bugaev:2013fya}
E.~Bugaev and P.~Klimai, ``{Axion inflation with gauge field production and
  primordial black holes},''
  \href{http://dx.doi.org/10.1103/PhysRevD.90.103501}{{\em Phys. Rev. D}
  {\bfseries 90} no.~10, (2014) 103501},
  \href{http://arxiv.org/abs/1312.7435}{{\ttfamily arXiv:1312.7435
  [astro-ph.CO]}}.

\bibitem{Cheng:2015oqa}
S.-L. Cheng, W.~Lee, and K.-W. Ng, ``{Numerical study of pseudoscalar inflation
  with an axion-gauge field coupling},''
  \href{http://dx.doi.org/10.1103/PhysRevD.93.063510}{{\em Phys. Rev. D}
  {\bfseries 93} no.~6, (2016) 063510},
  \href{http://arxiv.org/abs/1508.00251}{{\ttfamily arXiv:1508.00251
  [astro-ph.CO]}}.

\bibitem{Garcia-Bellido:2016dkw}
J.~Garcia-Bellido, M.~Peloso, and C.~Unal, ``{Gravitational waves at
  interferometer scales and primordial black holes in axion inflation},''
  \href{http://dx.doi.org/10.1088/1475-7516/2016/12/031}{{\em JCAP} {\bfseries
  12} (2016) 031}, \href{http://arxiv.org/abs/1610.03763}{{\ttfamily
  arXiv:1610.03763 [astro-ph.CO]}}.

\bibitem{Domcke:2017fix}
V.~Domcke, F.~Muia, M.~Pieroni, and L.~T. Witkowski, ``{PBH dark matter from
  axion inflation},''
  \href{http://dx.doi.org/10.1088/1475-7516/2017/07/048}{{\em JCAP} {\bfseries
  07} (2017) 048}, \href{http://arxiv.org/abs/1704.03464}{{\ttfamily
  arXiv:1704.03464 [astro-ph.CO]}}.

\bibitem{Garcia-Bellido:2017aan}
J.~Garcia-Bellido, M.~Peloso, and C.~Unal, ``{Gravitational Wave signatures of
  inflationary models from Primordial Black Hole Dark Matter},''
  \href{http://dx.doi.org/10.1088/1475-7516/2017/09/013}{{\em JCAP} {\bfseries
  09} (2017) 013}, \href{http://arxiv.org/abs/1707.02441}{{\ttfamily
  arXiv:1707.02441 [astro-ph.CO]}}.

\bibitem{Cheng:2018yyr}
S.-L. Cheng, W.~Lee, and K.-W. Ng, ``{Primordial black holes and associated
  gravitational waves in axion monodromy inflation},''
  \href{http://dx.doi.org/10.1088/1475-7516/2018/07/001}{{\em JCAP} {\bfseries
  07} (2018) 001}, \href{http://arxiv.org/abs/1801.09050}{{\ttfamily
  arXiv:1801.09050 [astro-ph.CO]}}.

\bibitem{Sorbo:2011rz}
L.~Sorbo, ``{Parity violation in the Cosmic Microwave Background from a
  pseudoscalar inflaton},''
  \href{http://dx.doi.org/10.1088/1475-7516/2011/06/003}{{\em JCAP} {\bfseries
  06} (2011) 003}, \href{http://arxiv.org/abs/1101.1525}{{\ttfamily
  arXiv:1101.1525 [astro-ph.CO]}}.

\bibitem{Cook:2011hg}
J.~L. Cook and L.~Sorbo, ``{Particle production during inflation and
  gravitational waves detectable by ground-based interferometers},''
  \href{http://dx.doi.org/10.1103/PhysRevD.86.069901,
  10.1103/PhysRevD.85.023534}{{\em Phys. Rev.} {\bfseries D85} (2012) 023534},
  \href{http://arxiv.org/abs/1109.0022}{{\ttfamily arXiv:1109.0022
  [astro-ph.CO]}}.
[Erratum: Phys. Rev.D86,069901(2012)].

\bibitem{Barnaby:2011qe}
N.~Barnaby, E.~Pajer, and M.~Peloso, ``{Gauge Field Production in Axion
  Inflation: Consequences for Monodromy, non-Gaussianity in the CMB, and
  Gravitational Waves at Interferometers},''
  \href{http://dx.doi.org/10.1103/PhysRevD.85.023525}{{\em Phys. Rev.}
  {\bfseries D85} (2012) 023525},
\href{http://arxiv.org/abs/1110.3327}{{\ttfamily arXiv:1110.3327
  [astro-ph.CO]}}.

\bibitem{Barnaby:2011vw}
N.~Barnaby, R.~Namba, and M.~Peloso, ``{Phenomenology of a Pseudo-Scalar
  Inflaton: Naturally Large Nongaussianity},''
  \href{http://dx.doi.org/10.1088/1475-7516/2011/04/009}{{\em JCAP} {\bfseries
  1104} (2011) 009},
\href{http://arxiv.org/abs/1102.4333}{{\ttfamily arXiv:1102.4333
  [astro-ph.CO]}}.

\bibitem{Anber:2012du}
M.~M. Anber and L.~Sorbo, ``{Non-Gaussianities and chiral gravitational waves
  in natural steep inflation},''
  \href{http://dx.doi.org/10.1103/PhysRevD.85.123537}{{\em Phys. Rev.}
  {\bfseries D85} (2012) 123537},
\href{http://arxiv.org/abs/1203.5849}{{\ttfamily arXiv:1203.5849
  [astro-ph.CO]}}.

\bibitem{Domcke:2016bkh}
V.~Domcke, M.~Pieroni, and P.~Binétruy, ``{Primordial gravitational waves for
  universality classes of pseudoscalar inflation},''
  \href{http://dx.doi.org/10.1088/1475-7516/2016/06/031}{{\em JCAP} {\bfseries
  1606} (2016) 031},
\href{http://arxiv.org/abs/1603.01287}{{\ttfamily arXiv:1603.01287
  [astro-ph.CO]}}.

\bibitem{Garcia-Bellido:2023ser}
J.~Garcia-Bellido, A.~Papageorgiou, M.~Peloso, and L.~Sorbo, ``{A flashing
  beacon in axion inflation: recurring bursts of gravitational waves in the
  strong backreaction regime},''
  \href{http://arxiv.org/abs/2303.13425}{{\ttfamily arXiv:2303.13425
  [astro-ph.CO]}}.

\bibitem{Garretson:1992vt}
W.~D. Garretson, G.~B. Field, and S.~M. Carroll, ``{Primordial magnetic fields
  from pseudoGoldstone bosons},''
  \href{http://dx.doi.org/10.1103/PhysRevD.46.5346}{{\em Phys. Rev. D}
  {\bfseries 46} (1992) 5346--5351},
  \href{http://arxiv.org/abs/hep-ph/9209238}{{\ttfamily arXiv:hep-ph/9209238}}.

\bibitem{Anber:2006xt}
M.~M. Anber and L.~Sorbo, ``{N-flationary magnetic fields},''
  \href{http://dx.doi.org/10.1088/1475-7516/2006/10/018}{{\em JCAP} {\bfseries
  10} (2006) 018}, \href{http://arxiv.org/abs/astro-ph/0606534}{{\ttfamily
  arXiv:astro-ph/0606534}}.

\bibitem{Caprini:2014mja}
C.~Caprini and L.~Sorbo, ``{Adding helicity to inflationary magnetogenesis},''
  \href{http://dx.doi.org/10.1088/1475-7516/2014/10/056}{{\em JCAP} {\bfseries
  10} (2014) 056}, \href{http://arxiv.org/abs/1407.2809}{{\ttfamily
  arXiv:1407.2809 [astro-ph.CO]}}.

\bibitem{Adshead:2016iae}
P.~Adshead, J.~T. Giblin, T.~R. Scully, and E.~I. Sfakianakis,
  ``{Magnetogenesis from axion inflation},''
  \href{http://dx.doi.org/10.1088/1475-7516/2016/10/039}{{\em JCAP} {\bfseries
  10} (2016) 039}, \href{http://arxiv.org/abs/1606.08474}{{\ttfamily
  arXiv:1606.08474 [astro-ph.CO]}}.

\bibitem{Jimenez:2017cdr}
D.~Jim\'enez, K.~Kamada, K.~Schmitz, and X.-J. Xu, ``{Baryon asymmetry and
  gravitational waves from pseudoscalar inflation},''
  \href{http://dx.doi.org/10.1088/1475-7516/2017/12/011}{{\em JCAP} {\bfseries
  12} (2017) 011}, \href{http://arxiv.org/abs/1707.07943}{{\ttfamily
  arXiv:1707.07943 [hep-ph]}}.

\bibitem{Durrer:2023rhc}
R.~Durrer, O.~Sobol, and S.~Vilchinskii, ``{Backreaction from gauge fields
  produced during inflation},''
  \href{http://dx.doi.org/10.1103/PhysRevD.108.043540}{{\em Phys. Rev. D}
  {\bfseries 108} no.~4, (2023) 043540},
  \href{http://arxiv.org/abs/2303.04583}{{\ttfamily arXiv:2303.04583 [gr-qc]}}.

\bibitem{Anber:2015yca}
M.~M. Anber and E.~Sabancilar, ``{Hypermagnetic Fields and Baryon Asymmetry
  from Pseudoscalar Inflation},''
  \href{http://dx.doi.org/10.1103/PhysRevD.92.101501}{{\em Phys. Rev. D}
  {\bfseries 92} no.~10, (2015) 101501},
  \href{http://arxiv.org/abs/1507.00744}{{\ttfamily arXiv:1507.00744
  [hep-th]}}.

\bibitem{Domcke:2019mnd}
V.~Domcke, B.~von Harling, E.~Morgante, and K.~Mukaida, ``{Baryogenesis from
  axion inflation},''
  \href{http://dx.doi.org/10.1088/1475-7516/2019/10/032}{{\em JCAP} {\bfseries
  10} (2019) 032}, \href{http://arxiv.org/abs/1905.13318}{{\ttfamily
  arXiv:1905.13318 [hep-ph]}}.

\bibitem{Domcke:2022kfs}
V.~Domcke, K.~Kamada, K.~Mukaida, K.~Schmitz, and M.~Yamada, ``{Wash-in
  leptogenesis after axion inflation},''
  \href{http://dx.doi.org/10.1007/JHEP01(2023)053}{{\em JHEP} {\bfseries 01}
  (2023) 053}, \href{http://arxiv.org/abs/2210.06412}{{\ttfamily
  arXiv:2210.06412 [hep-ph]}}.

\bibitem{Domcke:2018eki}
V.~Domcke and K.~Mukaida, ``{Gauge Field and Fermion Production during Axion
  Inflation},'' \href{http://dx.doi.org/10.1088/1475-7516/2018/11/020}{{\em
  JCAP} {\bfseries 11} (2018) 020},
  \href{http://arxiv.org/abs/1806.08769}{{\ttfamily arXiv:1806.08769
  [hep-ph]}}.

\bibitem{Domcke:2019qmm}
V.~Domcke, Y.~Ema, and K.~Mukaida, ``{Chiral Anomaly, Schwinger Effect,
  Euler-Heisenberg Lagrangian, and application to axion inflation},''
  \href{http://dx.doi.org/10.1007/JHEP02(2020)055}{{\em JHEP} {\bfseries 02}
  (2020) 055}, \href{http://arxiv.org/abs/1910.01205}{{\ttfamily
  arXiv:1910.01205 [hep-ph]}}.

\bibitem{Gorbar:2021rlt}
E.~V. Gorbar, K.~Schmitz, O.~O. Sobol, and S.~I. Vilchinskii, ``{Gauge-field
  production during axion inflation in the gradient expansion formalism},''
  \href{http://dx.doi.org/10.1103/PhysRevD.104.123504}{{\em Phys. Rev. D}
  {\bfseries 104} no.~12, (2021) 123504},
  \href{http://arxiv.org/abs/2109.01651}{{\ttfamily arXiv:2109.01651
  [hep-ph]}}.

\bibitem{DallAgata:2019yrr}
G.~Dall'Agata, S.~Gonz\'alez-Mart\'\i{}n, A.~Papageorgiou, and M.~Peloso,
  ``{Warm dark energy},''
  \href{http://dx.doi.org/10.1088/1475-7516/2020/08/032}{{\em JCAP} {\bfseries
  08} (2020) 032}, \href{http://arxiv.org/abs/1912.09950}{{\ttfamily
  arXiv:1912.09950 [hep-th]}}.

\bibitem{Domcke:2020zez}
V.~Domcke, V.~Guidetti, Y.~Welling, and A.~Westphal, ``{Resonant backreaction
  in axion inflation},''
  \href{http://dx.doi.org/10.1088/1475-7516/2020/09/009}{{\em JCAP} {\bfseries
  09} (2020) 009}, \href{http://arxiv.org/abs/2002.02952}{{\ttfamily
  arXiv:2002.02952 [astro-ph.CO]}}.

\bibitem{Peloso:2022ovc}
M.~Peloso and L.~Sorbo, ``{Instability in axion inflation with strong
  backreaction from gauge modes},''
  \href{http://dx.doi.org/10.1088/1475-7516/2023/01/038}{{\em JCAP} {\bfseries
  01} (2023) 038}, \href{http://arxiv.org/abs/2209.08131}{{\ttfamily
  arXiv:2209.08131 [astro-ph.CO]}}.

\bibitem{vonEckardstein:2023gwk}
R.~von Eckardstein, M.~Peloso, K.~Schmitz, O.~Sobol, and L.~Sorbo, ``{Axion
  inflation in the strong-backreaction regime: decay of the Anber-Sorbo
  solution},'' \href{http://arxiv.org/abs/2309.04254}{{\ttfamily
  arXiv:2309.04254 [hep-ph]}}.

\bibitem{Figueroa:2023oxc}
D.~G. Figueroa, J.~Lizarraga, A.~Urio, and J.~Urrestilla, ``{The strong
  backreaction regime in axion inflation},''
  \href{http://arxiv.org/abs/2303.17436}{{\ttfamily arXiv:2303.17436
  [astro-ph.CO]}}.

\bibitem{Figueroa:2020rrl}
D.~G. Figueroa, A.~Florio, F.~Torrenti, and W.~Valkenburg, ``{The art of
  simulating the early Universe -- Part I},''
  \href{http://dx.doi.org/10.1088/1475-7516/2021/04/035}{{\em JCAP} {\bfseries
  04} (2021) 035}, \href{http://arxiv.org/abs/2006.15122}{{\ttfamily
  arXiv:2006.15122 [astro-ph.CO]}}.

\bibitem{Figueroa:2021yhd}
D.~G. Figueroa, A.~Florio, F.~Torrenti, and W.~Valkenburg, ``{CosmoLattice: A
  modern code for lattice simulations of scalar and gauge field dynamics in an
  expanding universe},''
  \href{http://dx.doi.org/10.1016/j.cpc.2022.108586}{{\em Comput. Phys.
  Commun.} {\bfseries 283} (2023) 108586},
  \href{http://arxiv.org/abs/2102.01031}{{\ttfamily arXiv:2102.01031
  [astro-ph.CO]}}.

\bibitem{Gorbar:2021zlr}
E.~V. Gorbar, K.~Schmitz, O.~O. Sobol, and S.~I. Vilchinskii,
  ``{Hypermagnetogenesis from axion inflation: Model-independent estimates},''
  \href{http://dx.doi.org/10.1103/PhysRevD.105.043530}{{\em Phys. Rev. D}
  {\bfseries 105} no.~4, (2022) 043530},
  \href{http://arxiv.org/abs/2111.04712}{{\ttfamily arXiv:2111.04712
  [hep-ph]}}.

\bibitem{Sobol:2019xls}
O.~O. Sobol, E.~V. Gorbar, and S.~I. Vilchinskii, ``{Backreaction of
  electromagnetic fields and the Schwinger effect in pseudoscalar inflation
  magnetogenesis},'' \href{http://dx.doi.org/10.1103/PhysRevD.100.063523}{{\em
  Phys. Rev. D} {\bfseries 100} no.~6, (2019) 063523},
  \href{http://arxiv.org/abs/1907.10443}{{\ttfamily arXiv:1907.10443
  [astro-ph.CO]}}.

\bibitem{Sobol:2020lec}
O.~O. Sobol, A.~V. Lysenko, E.~V. Gorbar, and S.~I. Vilchinskii, ``{Gradient
  expansion formalism for magnetogenesis in the kinetic coupling model},''
  \href{http://dx.doi.org/10.1103/PhysRevD.102.123512}{{\em Phys. Rev. D}
  {\bfseries 102} no.~12, (2020) 123512},
  \href{http://arxiv.org/abs/2010.13587}{{\ttfamily arXiv:2010.13587
  [astro-ph.CO]}}.

\bibitem{Adshead:2019lbr}
P.~Adshead, J.~T. Giblin, M.~Pieroni, and Z.~J. Weiner, ``{Constraining axion
  inflation with gravitational waves from preheating},''
  \href{http://dx.doi.org/10.1103/PhysRevD.101.083534}{{\em Phys. Rev. D}
  {\bfseries 101} no.~8, (2020) 083534},
  \href{http://arxiv.org/abs/1909.12842}{{\ttfamily arXiv:1909.12842
  [astro-ph.CO]}}.

\bibitem{Adshead:2019igv}
P.~Adshead, J.~T. Giblin, M.~Pieroni, and Z.~J. Weiner, ``{Constraining Axion
  Inflation with Gravitational Waves across 29 Decades in Frequency},''
  \href{http://dx.doi.org/10.1103/PhysRevLett.124.171301}{{\em Phys. Rev.
  Lett.} {\bfseries 124} no.~17, (2020) 171301},
  \href{http://arxiv.org/abs/1909.12843}{{\ttfamily arXiv:1909.12843
  [astro-ph.CO]}}.

\end{thebibliography}\endgroup
  
\end{document}